\documentclass[onecolumn,aps,prd,groupedaddress,nofootinbib,showpacs]{revtex4}
\usepackage{graphicx}
\usepackage{amsmath}
\usepackage{bm}
\usepackage{slashed}
\usepackage{epsfig}
\usepackage{amsfonts}
\usepackage{color}
\usepackage{epstopdf}
\usepackage[pdftex]{hyperref}

\newcommand{\non}{\nonumber\\}

\begin{document}


\title{$\psi(2S)$ versus $J/\psi$ suppression in proton-nucleus collisions from factorization violating soft color exchanges}


\author{Yan-Qing Ma$^{1,2,3}$}
\author{Raju Venugopalan$^{4}$}
\author{Kazuhiro Watanabe$^{5,6,7}$}
\author{Hong-Fei Zhang$^{8}$}
\affiliation{
$^{1}$School of Physics and State Key Laboratory of Nuclear Physics and
Technology, Peking University, Beijing 100871, China\\
$^{2}$Center for High Energy Physics,
Peking University, Beijing 100871, China\\
$^{3}$Collaborative Innovation Center of Quantum Matter,
Beijing 100871, China\\
$^{4}$ Physics Department, Brookhaven National Laboratory, Upton, New York 11973-5000, USA. \\
$^{5}$ Key Laboratory of Quark and Lepton Physics (MOE) and Institute of Particle Physics, Central China Normal University, Wuhan 430079, China\\
$^{6}$ Physics Department, Old Dominion University, Norfolk, Virginia 23529, USA\\
$^{7}$ Theory Center, Jefferson  Laboratory, Newport News, Virginia 23606, USA\\
$^{8}$ Department of Physics, School of Biomedical Engineering, Third Military Medical University, Chongqing 400038, China.
}%
\date{\today}

\begin{abstract}
We argue that the large suppression of the $\psi(2S)$ inclusive cross-section relative to the $J/\psi$ inclusive cross-section in proton-nucleus (p+A) collisions can be attributed to factorization breaking effects in the formation of quarkonium. These factorization breaking effects arise from soft color exchanges between charm-anticharm pairs undergoing hadronization and comoving partons that are long-lived on time scales of quarkonium formation. We compute the short distance pair production of heavy quarks in the Color Glass Condensate (CGC) effective field theory and employ an improved Color Evaporation Model (ICEM) to describe their hadronization into quarkonium at large distances. The combined CGC+ICEM model provides a quantitative description of  $J/\psi$ and $\psi(2S)$ data  in proton-proton (p+p) collisions from both RHIC and the LHC. Factorization breaking effects in hadronization, due to additional parton comovers in the nucleus, are introduced heuristically by imposing a cutoff $\Lambda$, representing the average momentum kick from soft color exchanges, in the ICEM. Such soft exchanges have no perceptible effect on $J/\psi$ suppression in p+A collisions. In contrast, the interplay of the physics of these soft exchanges at large distances, with the physics of semi-hard rescattering at short distances, causes a significant additional suppression of $\psi(2S)$ yields relative to that of the $J/\psi$. A good fit of all RHIC and LHC $J/\psi$ and $\psi(2S)$ data, for transverse momenta $P_\perp\leq 5$ GeV in p+p and p+A collisions, is obtained for $\Lambda\sim 10$ MeV.

\end{abstract}

\pacs{11.80.La, 12.38.Bx,  14.40.Pq}

\maketitle

\section{Introduction}

Heavy quarkonium (Onium) production provides an important testing ground for the properties of strong interacting matter in Quantum Chromodynamics (QCD). A rigorous QCD  framework for Onium production is that of nonrelativistic QCD factorization (NRQCD) \cite{Bodwin:1994jh}. Within the NRQCD framework, many properties of Onium production in nucleon-nucleon collisions are now understood qualitatively, thanks to next-to-leading order (NLO) calculations of the short distance matrix elements~\cite{Ma:2010vd,Ma:2010yw,Butenschoen:2010rq,Gong:2013qka,Butenschoen:2012px,Chao:2012iv,Gong:2012ug}. However, there are still outstanding issues in the  application of NRQCD to world data on Onium production. Some of these are addressed in a recently proposed soft gluon factorization (SGF) \cite{Ma:2017xno} approach, which exhibits a much better convergence in the nonrelativistic velocity expansion relative to the NRQCD approach.

An improved treatment of Onium production is also feasible in the treatment of the short distance matrix elements in the kinematic regimes where higher twist and small-$x$ contributions are important. This is addressed within the framework of the Color Glass Condensate (CGC) effective theory \cite{Iancu:2003xm,Gelis:2010nm}, which  provides a systematic framework to account for the logs in $x$, as well as the higher twist contributions, that give rise to gluon saturation~\cite{Gribov:1984tu,Mueller:1985wy,McLerran:1993ni,McLerran:1993ka}.  Such a CGC+NRQCD framework~\cite{Kang:2013hta} provides a quantitative description of $J/\psi$ production in proton-proton (p+p) collisions~\cite{Ma:2014mri} and in proton-nucleus (p+A) collisions~\cite{Ma:2015sia,Qiu:2013qka}. The results can be matched  at large transverse momenta to the description of p+p and p+A collisions in the NLO pQCD+NRQCD framework~\cite{Ma:2010vd,Ma:2010yw,Chao:2012iv,Gong:2012ug}.
These comparisons to the data from RHIC and the LHC demonstrated that the color-octet contribution is nearly an order of magnitude larger than the color-singlet contribution, even at not too large transverse momenta. Therefore,
reasonable results can be obtained by applying the simpler Color Evaporation
model (CEM)~\cite{Fritzsch:1977ay,Gluck:1977zm,Barger:1979js} of Onium formation, which mainly includes the contribution of color-octet configurations.  Prior studies of $J/\psi$ production in p+A collisions within the CGC+CEM can be found in Refs. \cite{Fujii:2006ab,Fujii:2013gxa,Ducloue:2015gfa,Watanabe:2015yca}. In Ref.~\cite{Ma:2016exq}, an improved Color Evaporation Model (ICEM) was introduced, which took into account the kinematic constraints relating the momentum of the charm pair to that of the produced Onium. This improved treatment of the kinematics helps explain the transverse momentum $P_\perp$ dependence of data on the ratio of the $\psi(2S)$ to $J/\psi$ yields, which are independent of $P_\perp$ in the CEM.

Recently, the PHENIX Collaboration at RHIC reported on measurements of $\psi(2S)$ production in d+Au collisions with center-of-mass energy $\sqrt{s_{NN}}=0.2$ TeV/nucleon~\cite{Adare:2013ezl}. They found that in rare events corresponding to a large number of collisions, the $\psi(2S)$ yield is significantly suppressed relative to p+p collisions. This suppression is greater than that seen for the $J/\psi$ yield. The observation of $\psi(2S)$ suppression was corroborated by the ALICE Collaboration in p+Pb collisions at $\sqrt{s_{NN}}=5.02$ TeV/nucleon~\cite{Abelev:2014zpa}. They found that the suppression parameter for $\psi(2S)$, $R_{\rm pA}^{\psi(2S)}$, is smaller than $0.6$, even for $P_\perp$ as large as 7 GeV at rapidities towards the proton fragmentation region. The $\psi(2S)/J/\psi$ suppression in p+A collisions is also seen by the LHCb Collaboration~\cite{Aaij:2016eyl} and in more detailed  studies by both the ALICE Collaboration~\cite{Adam:2016ohd} and the PHENIX Collaboration~\cite{Adare:2016psx}. Key features of the experimental results are (i) the ratio $R_{\rm pA}^{\psi(2S)}/R_{\rm pA}^{J/\psi}$ decreases nearly linearly with increasing number of produced charged particles $N_{\text{ch}}$; (ii) $R_{\rm pA}^{\psi(2S)}$ likewise decreases with the increasing number of collisions $N_{\text{coll}}$; (iii) $R_{\rm pA}^{\psi(2S)}<0.6$ at the LHC for both forward rapidity and backward rapidities; and (iv) $R_{\rm pA}^{\psi(2S)}$ is nearly flat at 0.6 for forward rapidities even as $P_\perp$ becomes larger while at backward rapidities, it goes to unity with increasing $P_\perp$.

\begin{figure}
	\centering
	\includegraphics[width=0.5\linewidth]{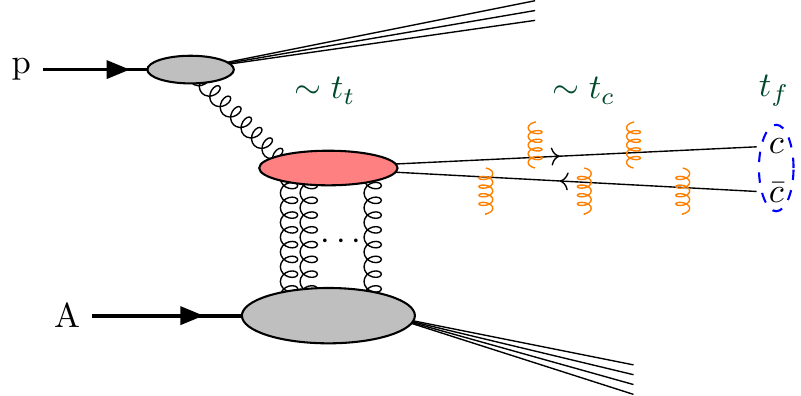}
	\caption{(Color online)
		Schematic diagram of Onium production in p+A collisions. The red blob represents parton hard scattering  at short distances. Vertical gluons represent multiple gluon scattering  (with typical net momentum exchange of order $Q_s$, the saturation scale) off the target nucleus, expressed through a lightlike Wilson line. Soft gluon exchange between produced $c\bar c$ pair and comover spectators at larger distances are shown as orange vertical gluons. The three time scales ($t_t$, $t_c$, $t_f$) discussed in the text are also illustrated in the figure.
	}
	\label{fig:diagram}
\end{figure}

To arrive at a deeper understanding of the systematics of these striking results, it is useful to first consider the different time scales that are relevant for Onium production in p+A collisions. Proceeding from short to long time scales (or distances) at collider energies, the first is the time scale for the $c\bar{c}$ pair to traverse the nucleus ($t_t$). The second is the time scale of $c\bar{c}$ pair production ($t_c$), and the last is the time scale for Onium formation ($t_f$). If the Onium $\psi$ is produced in the forward rapidity region, these time scales in the laboratory frame are given by
\begin{eqnarray}
t_t&\sim& 2 R_A\frac{m_n}{E_n}, \\
t_c&\sim& \frac{1}{2m}\,\frac{E}{m}>\frac{1}{2m}, \\
t_f&\sim& \frac{1}{mv^2}\,\frac{E}{m}\sim\frac{t_c}{v^2},
\end{eqnarray}
where $m_n$ is the mass of the proton and $E_n$ is the energy of the nucleus per nucleon in the rest frame of the proton, while likewise $m$ and $E$ denote the mass and the energy of the Onium state.
Further, $v^2\approx 0.3$ is the square of the relative velocity of the charm quarks in the $\psi$ rest frame, and $R_A$ is the radius of the nucleus. The value of $R_A$ is estimated to be about $5\,\mathrm{fm}$ in Ref.~\cite{Ma:2015sia}, which implies $t_t\sim 0.05$ fm for PHENIX and $t_t\sim 0.002$ fm for ALICE. Considering that $t_c>0.07$ fm for charm quarks, we find that the hierarchy of time scales $t_t<t_c<t_f$ is satisfied at both RHIC and the LHC (Fig.~\ref{fig:diagram}). These simple considerations suggest that models that explain the suppression of Onium yields at lower energies as occuring due to nuclear absorption effects~\cite{Arleo:1999af} are implausible at higher energies: this is because the Onia are formed well outside the nucleus.

Several theoretical works have since addressed this unanticipated result of $\psi(2S)/J/\psi$ suppression in p+A collisions. In Ref.~\cite{Ferreiro:2014bia}, this suppression was explained as occurring due to the interaction of the $J/\psi$ and $\psi(2S)$ mesons with comovers. The latter, as the term suggests, are hadrons that travel along with the $c\bar c$ pair and scatter off it, dissociating the lightly bound $\psi(2S)$ more easily than the $J/\psi$. In Ref.~\cite{ Kisslinger:2014rya}, the author proposed that $\psi(2S)$ is a state with equal amounts of mixing between normal charmonium and hybrid charmonium, and it is the hybrid charmonium that suffers the larger suppression. In Ref.~\cite{Du:2015wha}, the authors implemented hadronic reaction rates into a thermal rate equation framework and introduced final state effects to explain the large suppression of $\psi(2S)$. In Ref.~\cite{Chen:2016dke}, the authors proposed that there are hot medium effects in additional to cold medium effects; while cold medium effects are similar for both the $J/\psi$ and $\psi(2S)$, the authors propose that hot medium effects are much more important for the $\psi(2S)$.

In this paper, we will argue that there is a hitherto little considered dynamical effect already at the parton level that is sufficient to explain the systematics of the data $\psi(2S)/J/\psi$ suppression. Before hadron comovers form, there are parton comovers which, due to time dilation, hadronize on longer time scales than the $c\bar{c}$ pair. These partons can have soft color exchanges with momenta of order or less than $\Lambda_{\rm QCD}$, the intrinsic QCD scale. For the $J/\psi$, such color exchanges have little effect on the suppression because the $J/\psi$ mass is well below the open charm threshold of $D\bar{D}$ pairs. In contrast, the $\psi(2S)$ mass is close to the $D\bar{D}$ threshold. Multiple scattering of the $c\bar{c}$ pair off the nucleus, modifies its mass spectrum, making it more susceptible to break-up, even with very soft color exchanges with average momentum $\Lambda \ll \Lambda_{\rm QCD}$. These soft exchanges represent factorization breaking in the fragmentation of different charmonium states.  The momentum scale for such exchanges, $\Lambda\sim10\!\!-\!\!20$\;MeV, is much smaller than the typical freeze-out temperature scales in heavy-ion collisions.

This paper is organized as follows. In the next section, we will flesh out the above argument and quantify it in the ICEM framework. In Sec.~III, we will recapitulate some of the essential details of the CGC  computation of the $c\bar{c}$ invariant mass cross-section. The results in the CGC+ICEM framework are compared in Sec.~IV to data from p+p and p+A collisions. Section~V summarizes our results and suggests further tests and refinements of the framework. In an Appendix, we present in tabular form, the values of the nonperturbative parameter $F_\psi$ in the ICEM, extracted from fits of $J/\psi$ and $\psi(2S)$  cross-sections to data in p+p collisions
for a range in quark masses, energies and rapidities.

\section{Factorization breaking and $\psi(2S)$ suppression}\label{sec2}

\begin{figure}
	\centering
	\includegraphics[width=0.6\linewidth]{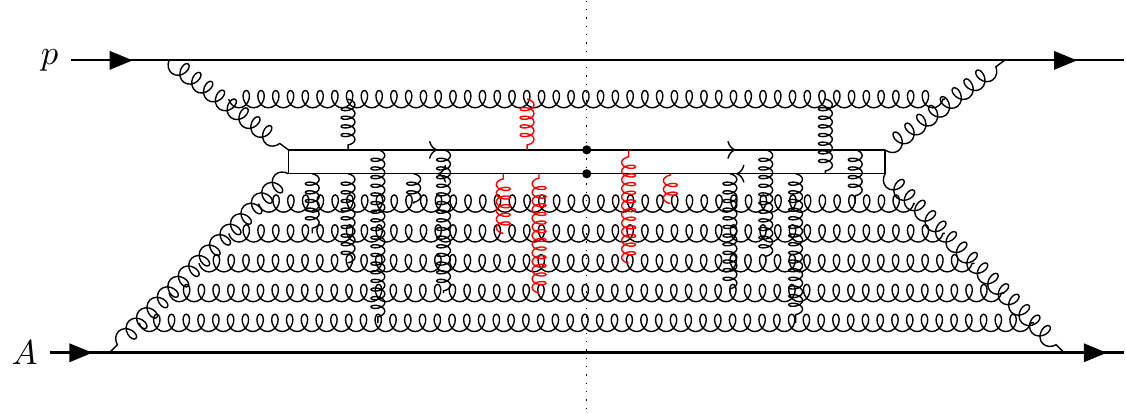}
	\caption{(Color online)
	An illustration of $c\bar c$ pair production  and hadronization in p+A collisions. The left side of the final state cut (represented by a dashed vertical line) is the scattering amplitude  while the right side is its complex conjugate. The blobs represent the charmonium final state. Gluon exchanges of the $c\bar{c}$ with the target, carrying momenta of order $Q_s$, are shown in black. Soft color exchanges with momentum resolution $\Lambda$ are shown in red. See text for further discussion.
	}
	\label{fig:diagram2}
\end{figure}

We will consider first the production of $c\bar{c}$ pairs within the dilute-dense framework of the CGC and subsequently the hadronization of these charm pairs in the ICEM framework. In the dilute-dense CGC framework, a gluon from the proton projectile emits the $c\bar{c}$ pair either before or after the proton scatters off the nuclear target.\footnote{Emissions from within the target are suppressed~\cite{Blaizot:2004wv} by the $\gamma$ factor corresponding to the Lorentz contracted width of the target.} This gluon in turn is emitted from color sources at higher rapidities, which are static sources over the lifetime of the gluon and the $c\bar{c}$ pair. Because there are several of these sources, their collective color charge lives in a higher dimensional representation of $SU(3)$; therefore their coupling to the process of interest can be represented by a classical color charge density $\rho_p$. Likewise, the color charge density of sources from the nuclear target that emits a gluon that scatters off the $c\bar{c}$ pair can be denoted by $\rho_A$.  The dilute-dense approximation\footnote{Here one is presuming that there exists a limit where the source lives in a classical representation even though it is dilute.}  corresponds to the $\rho_p/k_{1\perp}^2 \ll 1$ and $\rho_A/k_{2\perp}^2 \sim 1$.

This dilute-dense approximation is a powerful one and allows for the treatment of charmonium production that factorizes the contributions from the projectile and the target. The explicit expression for this factorized cross-section is given in the next section. It resums all semi-hard multiple scattering contributions from the target that are shown in Fig.~\ref{fig:diagram2}.  As we noted previously, this short distance framework can be matched to NRQCD at large distances after projecting the charmonium pair cross-section on to color-singlet and color-octet configurations. However, this factorization is by no means assured, and soft color exchanges between the comoving sources and the charm pair both before and after hadronization.

These soft color exchanges are depicted by the red vertical gluon lines in Fig.~\ref{fig:diagram2}. For the $J/\psi$ production cross-section, we will show that they have little impact. This is not the case for the $\psi(2S)$, which is much more weakly bound, and close to the $D\bar{D}$ threshold. In p+p collisions, in the CEM model~\cite{Fritzsch:1977ay,Gluck:1977zm,Barger:1979js}, the nonperturbative transition into the bound state is parametrized by a single parameter $F_\psi$ for each Onium state, here generically denoted by $\psi$. As we shall discuss, these are fit to data; thus, even though the $\psi(2S)/J/\psi$ ratio is much smaller than unity, the effect of soft color exchanges is indistinguishable from other nonperturbative effects that are all absorbed into $F_\psi$.

Studies of the $\psi(2S)/J/\psi$ cross-section in p+A collisions therefore provide an opportunity to investigate the role of these soft color exchanges. Firstly, we assume that all other nonperturbative effects are universal and therefore accounted for in $F_\psi$. Secondly, since there are more color sources in a nucleus, the role of soft color exchanges should not be universal, but should be $A$ dependent. As noted, their effect should not be visible for the $J/\psi$ cross-section, but may influence the $\psi(2S)$ cross-section. We will account for this effect heuristically by writing the Onium differential cross-section as
\begin{align}\label{eq:model}
\frac{d\sigma_\psi}{d^3\vec{P}}=F_{\psi}\int_{m_\psi}^{2m_D-\Lambda}dM\frac{d\sigma_{c\bar c}(M,\vec{P^\prime})}{dM d^3\vec{P}} \, .
\end{align}
Here, $d\sigma_{c\bar c}(M,\frac{M}{m_\psi}\vec{P})$ is the differential cross-section to produce a $c\bar c$ pair with an invariant mass $M$. This distribution is also a function of momentum $\vec{P^\prime}=\frac{M}{m_\psi} \vec{P}$. This multiplicative factor shifting the momentum from $\vec{P}\rightarrow \vec{P^\prime}$ is a key feature of the ICEM~\cite{Ma:2016exq}. Another important feature of ICEM is a new lower bound of the $M$-integral, which results in that the size of the $M$-integral range is close to the binding energy of $\psi$. Because the binding energy of the $\psi(2S)$ is smaller than that of the $J/\psi$, the aforementioned soft color exchanges should have greater effect for $\psi(2S)$ production. As discussed in Ref.~\cite{Ma:2016exq}, these two features  arise from careful power counting in relating the momentum of the $c\bar{c}$ pair to that of the produced Onium. As we shall discuss in the following section, the features in ICEM indeed enable us to describe correctly the ratio of $\psi(2S)$ to $J/\psi$. Further, $m_D$ is the mass of $D$ meson, and, as noted previously, $F_{\psi}$ is the transition probability governing the nonperturbative conversion of $c\bar c$ to $\psi$. The enhancement of soft color exchanges in nuclei is represented by the soft scale $\Lambda$, which appears in the upper limit of the integration of $M$. It quantifies the additional kick given by nuclear parton comovers, over and above the soft color exchange effects, whose kinematic effects are incorporated in the ICEM. We will quantify these ideas further in the next section.

\section{Production of $c\bar c$ pair}{\label{section:ccbar}}

We will briefly outline here the CGC computation for the process ${\rm p}+A\rightarrow c\;(p) + \bar c\;(q)+X$~\cite{Blaizot:2004wv,Fujii:2006ab},
where $(p)$ and $(q)$ respectively represent the momenta of the $c$ and $\bar c$. The $c\bar c$ pair is produced with the total transverse momentum  $P_\perp=p_{\perp}+q_{\perp}$ at the rapidity  $y=\frac{1}{2}\ln\left(\frac{p^++q^+}{p^-+q^-}\right)$. The longitudinal momentum fractions of the projectile proton and target nucleus carried by incoming gluons are represented by
\begin{align}
x_{1,2}=\sqrt{\frac{M^2+P_\perp^2}{s}}e^{\pm y}\,,
\label{eq:lo-kinematics}
\end{align}
where $M$ is the invariant mass of the $c\bar c$ and $s$ is the center-of-mass energy per nucleon of the p+A collision.
The leading order pair production cross-section for this process can be expressed as
\begin{align}
\frac{d \sigma_{q \bar{q}}}{d^2p_{\perp} d^2q_{\perp} dy_p dy_{q}}
=
\frac{\alpha_s^2}{64\pi^6 C_F}
\int\frac{d^2k_{2\perp}d^2k_\perp}{(2\pi)^4}
\frac{\Xi({k}_{1\perp}, {k}_{2\perp},{k}_{\perp})}
{k_{1\perp}^2 k_{2\perp}^2}
\;
\varphi_{{\rm p},x_1}(k_{1\perp})
\;
\phi_{A,x_2}({k}_{2\perp},{k}_\perp)\,.
\label{eq:xsection-kt-factorization-LN}
\end{align}
The hard scattering contribution  $\Xi$  can be decomposed into the individual pieces $\Xi=\Xi^{q\bar{q},q\bar{q}}+\Xi^{q\bar{q},g}+\Xi^{g,g}$, where
\begin{align}
&\Xi^{q\bar{q},q\bar{q}}
=\;\frac{32 p^+q^+(m^2+a_\perp^2)(m^2+b_\perp^2)}{[2p^+(m^2+a_\perp^2)+2q^+(m^2+b_\perp^2)]^2}\,,\non
&\Xi^{q\bar{q},g}
=\;\frac{16}{2(m^2+p\cdot q)[2p^+(m^2+a_\perp^2)+2q^+(m^2+b_\perp^2)]}
\Bigg[(m^2+a_\perp\cdot b_\perp)\left\{q^+C\cdot p+p^+C\cdot q-C^+(m^2+p\cdot q)\right\}\non
&+C^+\left\{(m^2+b_\perp\cdot q_\perp)(m^2-a_\perp\cdot p_\perp)-(m^2+a_\perp\cdot q_\perp)(m^2-b_\perp\cdot p_\perp)\right\}\non
&+p^+\left\{a_\perp\cdot C_\perp(m^2+b_\perp\cdot q_\perp)-b_\perp\cdot C_\perp(m^2+a_\perp\cdot q_\perp)\right\}
+q^+\left\{a_\perp\cdot C_\perp(m^2-b_\perp\cdot p_\perp)-b_\perp\cdot C_\perp(m^2-a_\perp\cdot p_\perp)\right\}\Bigg]\,,\non
&\Xi^{g,g}
=\;\frac{4\left[2(p\cdot C)(q\cdot C)-(m^2+p\cdot q)C^2\right]}{4(m^2+p\cdot q)^2}\,.
\end{align}
In the above, $a_\perp=q_\perp-k_\perp$ and $b_\perp=q_\perp-k_\perp-k_{1\perp}$. The Lipatov vertex~\cite{Lipatov:1996ts} $C^\mu$ that appears here, can be written in component form as $C^+=\;p^++q^+-\frac{k_{1\perp}^2}{p^-+q^-}$, $C^-=\;\frac{k_{2\perp}^2}{p^++q^+}-(p^-+q^-)$, and  $C_\perp=\;k_{2\perp}-k_{1\perp}$.

The unintegrated gluon distribution $\varphi_{{\rm p},x}(k_\perp)$ of the projectile proton depends explicitly on the transverse momentum of the gluon inside the proton, and can be expressed as
\begin{align}
\varphi_{{\rm p},x}(k_{\perp}) = \pi R_{\rm p}^2\, \frac{N_ck_{\perp}^2}{4\alpha_s} \widetilde{F}_{x}(k_{\perp})\,,
\label{eq:proton-wavefn}
\end{align}
where $\pi R_{\rm p}^2$ is the transverse area occupied by gluons in the proton and $\widetilde{F}_{x}(k_{\perp})$ is the Fourier transform of the dipole amplitude in the adjoint representation; in the large $N_c$ limit, this is simply the square of the fundamental dipole amplitude $S_x(x_\perp)$~\cite{Gelis:2006tb}. One therefore obtains
\begin{align}
\widetilde{F}_{x}(k_{\perp})
= \int d^2x_\perp e^{-ik_{\perp}\cdot x_\perp}
{S}^2_{x}(x_\perp)
=\int\frac{d^2l_\perp}{(2\pi)^2}F_{x}({k}_{\perp}-{l}_\perp)F_{x}(l_{\perp}) \,,
\label{eq:dipole-product}
\end{align}
with
\begin{align}
{F}_{x}(k_{\perp})
\equiv\int d^2x_\perp e^{-ik_{\perp}\cdot x_\perp} {S}_{x}(x_\perp)=\int d^2x_\perp e^{-ik_{\perp}\cdot x_\perp} \frac{1}{N_c}\left<{\mathrm Tr}\left[U(x_\perp)U^\dagger(0_\perp)\right]\right>_x \, .
\end{align}
The  $U(x_\perp)$ in the rightmost expression is the fundamental Wilson line representing multiple scattering of the quark with the background fields at the position  $x_\perp$ in the amplitude and $U^\dagger(0_\perp)$ is the corresponding Wilson line in the complex conjugate amplitude at the spatial position $0_\perp$.

The function $\phi_{A,x_2}({k}_{2\perp},{k}_\perp)$ in Eq.~(\ref{eq:xsection-kt-factorization-LN}) is a multi-point Wilson line correlator in the nuclear target. In the large-$N_c$ approximation, it can be expressed as
\begin{align}
\phi_{{A},x}({k}_{\perp},{l}_\perp)
=\pi R_{A}^2 \,\frac{N_c k_{\perp}^2}{4\alpha_s}F_{x}({k}_{\perp}-{l}_\perp)F_{x}(l_\perp)\,,
\end{align}
where $\pi R_{A}^2$ is effective transverse area of the nucleus. Therefore, both $\varphi_{{\rm p},x}$ and $\phi_{{A},x}$ can be  expressed in terms of the dipole amplitude in the fundamental representation. The rapidity (or energy) dependence of the differential cross section in Eq.~(\ref{eq:xsection-kt-factorization-LN}) for $c\bar c$ production is given entirely by the evolution of the dipole amplitude with rapidity.

In the CGC, the rapidity dependence of the dipole amplitude, to leading accuracy in $N_c$, is given by the  Balitsky-Kovchegov (BK) equation~\cite{Balitsky:1995ub,Kovchegov:1996ty}:
\begin{align}
-\frac{dS_{x}({r_\perp})}{dY}
 = \int d^2 r_{1\perp} \mathcal{K}_{\rm run}(r_\perp, r_{1\perp})
\Big [  S_{x}({r_\perp}) - S_{x}({r_{1\perp}})S_{x}({r_{2\perp}})
\Big ],
\end{align}
where $Y=\ln1/x$, and the  running coupling evolution kernel in Balitsky's prescription~\cite{Balitsky:2006wa} is given by
\begin{align}
\mathcal{K}_{\rm run}(r_\perp,r_{1\perp})=&
\frac{\alpha_s (r_\perp^2) N_c} {2\pi^2}\,
\left [
\frac{1}{r_{1\perp}^2} \left ( \frac{\alpha_s(r_{1\perp}^2)}{\alpha_s(r_{2\perp}^2)}-1  \right )
+
\frac{r_\perp^2}{r_{1\perp}^2 r_{2\perp}^2}
+
\frac{1}{r_{2\perp}^2} \left ( \frac{\alpha_s(r_{2\perp}^2)}{\alpha_s(r_{1\perp}^2)}-1  \right )
\right ],
\label{eq:rcBK-kernel}
\end{align}
where ${r}_\perp= {r}_{1\perp}+ {r}_{2\perp}$ is the size of the ``parent"  dipole size prior to one step in $Y$ evolution. The initial condition of the rcBK equation can be determined by a fit to the HERA-DIS data available below $x_0=0.01$. Although uncertainties with respect to the choice of the form of the initial condition for the rcBK equation remain, one can set the initial dipole amplitude at $x=x_0$ to be of the form given by the McLerran-Venugopalan (MV) model~\cite{McLerran:1993ni,McLerran:1993ka}:
\begin{align}
S_{{x=x_0}}(r_\perp)=
\exp\left[-\frac{\left(r_\perp^2Q_{s0,{\rm p}}^2\right)^\gamma}{4}\ln\left(\frac{1}{r_\perp\Lambda^\prime}+e\right)\right],
\label{eq:IC-rcBK}
\end{align}
where $\gamma$ is an anomalous dimension, $Q_{s0,{\rm p}}$ is the saturation scale in the proton at $x=x_0$, and the one loop coupling constant in coordinate space $\alpha_s(r_\perp^2)= \left [\frac{9}{4\pi} \ln \left (\frac{4 C^2}{r_\perp^2\Lambda^{\prime2}}+a \right ) \right ]^{-1}$ is employed to solve the rcBK equation. The parameters in this initial condition obtained from the fit to HERA data are given in Ref.~\cite{Albacete:2010sy}.

For our purposes, the MV model parametrization (with $\gamma=1$) is sufficient to describe the data on Onium production~\cite{Ma:2015sia,Ma:2014mri,Fujii:2013gxa}. For the initial input parameters, we will choose $Q_{s0,\rm p}^2=0.2\;{\rm GeV}^2$, $\Lambda^\prime=0.241\;{\rm GeV}$, $\gamma=1$, and $C=1$ as previous implemented in Ref.~\cite{Fujii:2013gxa}. The infrared cutoff $a$ is chosen to satisfy $\alpha_s(r \to \infty)=0.5$. For the target nucleus,  $Q_{s0,A}^2=cA^{1/3}Q_{s0,{\rm p}}^2$ where $c\lesssim0.5$ for minimum bias events in p+A collisions~\footnote{In Ref.~\cite{Dusling:2009ni}, a small value of $c\approx 0.25$ was shown to fit the New Muon Collaboration data on the nuclear structure functions $F_{2,A}(x,Q^2)$.}. Due to the significant uncertainties in these determinations, we shall vary  $Q_{s0,A}^2=(1.5\!\!-\!\!2.0)\,Q_{s0,{\rm p}}^2$ for heavy nuclei such as Pb and Au in our numerical computations.

At forward rapidities, values of $x\geq x_0$ are accessed in the proton wavefunction. We therefore need to extrapolate the parametrization of the dipole amplitude to these $x$ values. Following the discussion in Ref.~\cite{Ma:2014mri}, the adjoint dipole distribution in Eq.~(\ref{eq:proton-wavefn}) at $x\geq x_0$ can be determined to be
\begin{align}
\widetilde{F}_x(k_\perp)\overset{x> x_0}{=}\;a(x)\widetilde{F}_{x_0}(k_\perp)
\end{align}
where
\begin{align}
a(x)\overset{x> x_0}{=}  xG (x,Q_{0}^2)\,\left[\frac{\pi R_{\rm p}^2 N_c}{4\pi^3\,4\alpha_s} \int_0^{Q_{0}^2} dk_{1\perp}^2\,k_{1\perp}^2 \widetilde{F}_{x_0}(k_{1\perp})\right]^{-1} \, .
\end{align}

Requiring $a(x)=1$ and $a^\prime (x)=0$ at $x=x_0$, is sufficient to determine both $R_{\rm p}$ and $Q_0$ simultaneously. Indeed, utilizing the CTEQ6M parton distribution set~\cite{Pumplin:2002vw} for $xG(x,Q_0^2)$ and two loop strong coupling constant with $n_f=4$ and $\Lambda=326$\;MeV gives $Q_0=8.10$\;GeV and $R_{\rm p}=0.438$\;fm.

\section{Numerical results}
We will begin this section by first discussing some features of the ICEM distributions that are employed in our fits to p+p and p+A data on Onium production.
We will then make quantitative comparisons to p+p data at RHIC and LHC energies, and subsequently to data from p+A collisions at both colliders.
\subsection{Remarks on the ICEM}

\begin{figure}
	\centering
	\includegraphics[width=0.45\linewidth]{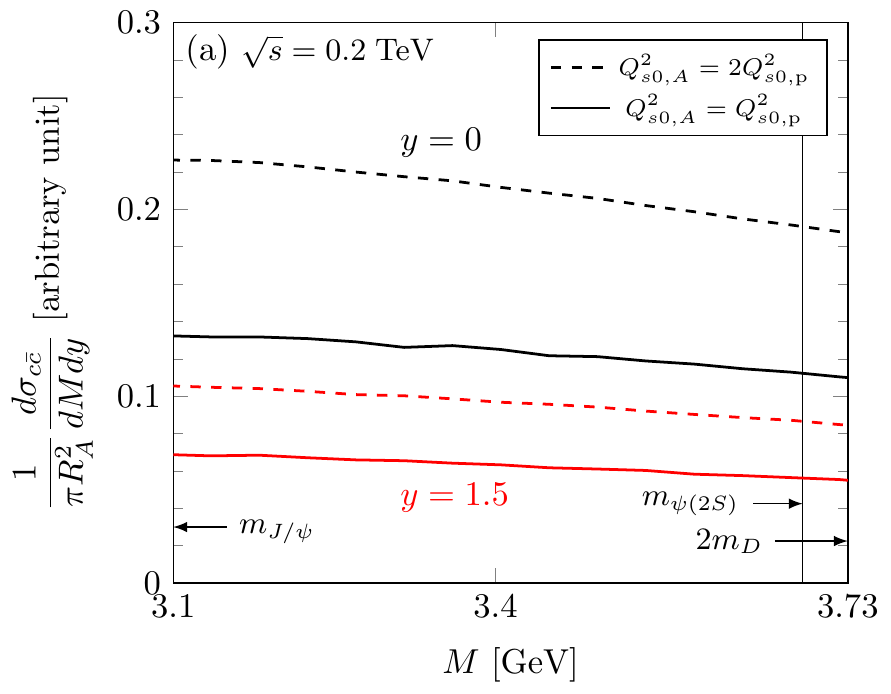}
	\includegraphics[width=0.45\linewidth]{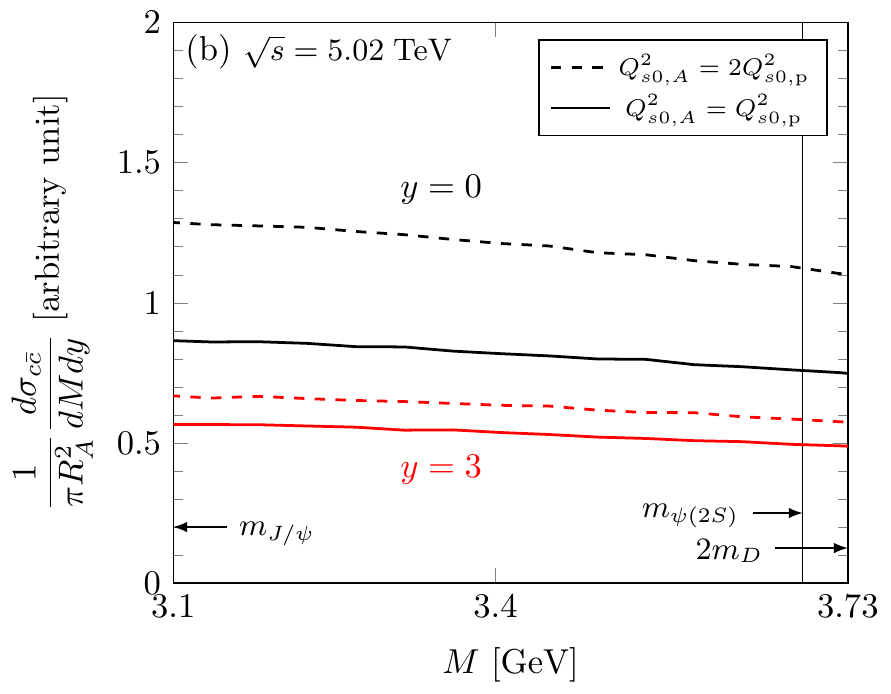}
	\caption{(Color online)
	$M$ distribution of the $c\bar c$ pair production cross-section for different rapidities and initial saturation scales in the target nucleus. The  figure (a) [(b)] is the result at a RHIC (LHC) energy with $m=1.3$\;GeV. The boundaries of the $M$-integral for  ($m_{J/\psi}$, $m_{\psi(2S)}$, $2m_D$) are specified.
	}
	\label{fig:M-distribution}
\end{figure}

As noted, the differential cross section for  $c\bar c$ production in p+A collisions is calculated using Eq.~(\ref{eq:xsection-kt-factorization-LN}). We will also assume in this paper that Eq.~(\ref{eq:xsection-kt-factorization-LN}) is applicable to p+p collisions, as was also assumed previously in Ref.~\cite{Ma:2014mri}. The expression in the  ICEM  of Eq.~(\ref{eq:model}) can be reexpressed as
\begin{align}
\frac{d\sigma_\psi}{d^2P_\perp dy}=F_{\psi}\int_{m_\psi}^{2m_D-\Lambda}dM \left(\frac{M}{m_\psi}\right)^2\frac{d\sigma_{c\bar c}}{dM d^2P^\prime_\perp dy}\Bigg|_{P^\prime_\perp=\frac{M}{m_\psi}P_\perp},
\end{align}
where $m_{\psi}=3.1\;{\rm GeV}$ for $J/\psi$ production, $m_{\psi}=3.686\;{\rm GeV}$ for $\psi(2S)$ production, and $2\,m_D=3.728\;{\rm GeV}$.
The transition probability $F_{\psi}$ includes the $K$-factor incorporating higher order corrections as well as feed down contributions from excited states. We choose $F_{\psi}$ to fit data by minimizing $\chi^2$.  As mentioned in Sec.~\ref{sec2}, we introduce a cutoff $\Lambda$ to parametrize the enhancement of soft color exchanges in p+A collisions. In our computations, we will therefore set $\Lambda=0$ for p+p collisions and vary it for p+A collisions. In addition, we do not, for simplicity, consider the possible $P_\perp$ and rapidity dependence of $\Lambda$.
Since the Onium production cross-section is a leading order result, we will choose the strong coupling constant in Eq.~(\ref{eq:xsection-kt-factorization-LN}) to be $\alpha_s(Q_0)$ (where as stated above, $Q_0=8.1$ GeV) throughout in our numerical computations.

In Fig.~\ref{fig:M-distribution}, we show the invariant mass ($M$) distribution of the $c\bar c$ pair production in the CGC framework at RHIC and LHC energies obtained by varying the rapidity, quark mass, and the initial saturation scale for the target nucleus. We observe that the yields in $c\bar c$ pair production increase as the initial saturation scale increases, although the enhancements are smaller at forward rapidity ($y=4$) because the results are sensitive to large $x$ distributions in the proton.
Of particular importance for the $M$ distributions is the fact that the phase space of the produced $c\bar c$ pair is limited to lie within the narrow range between $m_{\psi(2S)}$ and $2m_{D}$ when the $c\bar c$ pair is transformed into $\psi(2S)$. For $J/\psi$ production, the $c\bar c$ pair has a significantly  larger phase space than that for $\psi(2S)$ production. Therefore, introducing the cutoff $\Lambda$ can affect $\psi(2S)$ production.

\subsection{Results for p+p collisions}

\begin{figure}
	\centering
	\includegraphics[width=0.7\linewidth]{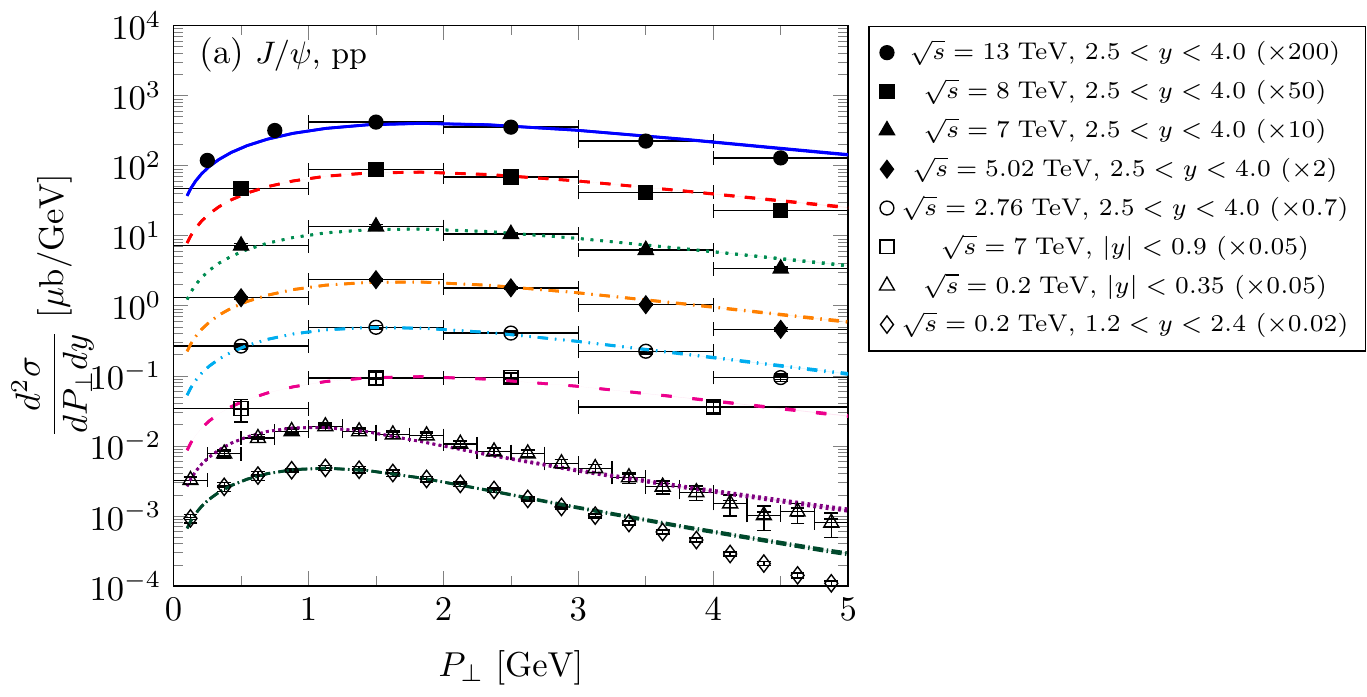}
	\includegraphics[width=0.7\linewidth]{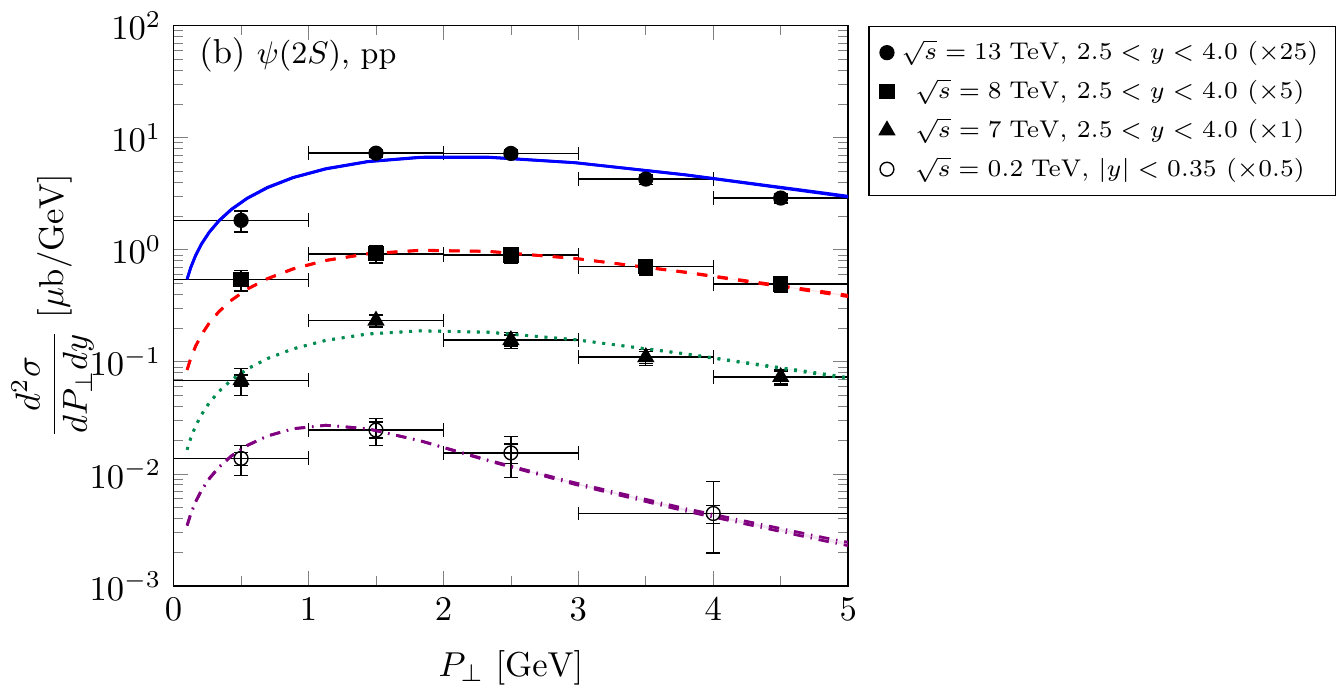}
	\caption{(Color online)
	Differential cross-section as a function of $P_\perp$ for $J/\psi$ and $\psi(2S)$ production in p+p collisions at RHIC and the LHC in the CGC+ICEM framework. The various lines correspond to the results for the different values of $\sqrt{s}$ or $y$-range. The uncertainty bands reflect the quark mass dependence: $m=(1.3\!\!-\!\!1.4)$\;GeV, though the width of the bands is narrow. Data are taken from Refs.~\cite{Adare:2011vq,Aamodt:2011gj,Abelev:2012kr,Abelev:2014qha,Adam:2015rta,Acharya:2017hjh}.
	}
	\label{fig:Jpsi-Psi2S-pt-distribution-pp}
\end{figure}

\begin{figure}
	\centering	
	\includegraphics[width=0.45\linewidth]{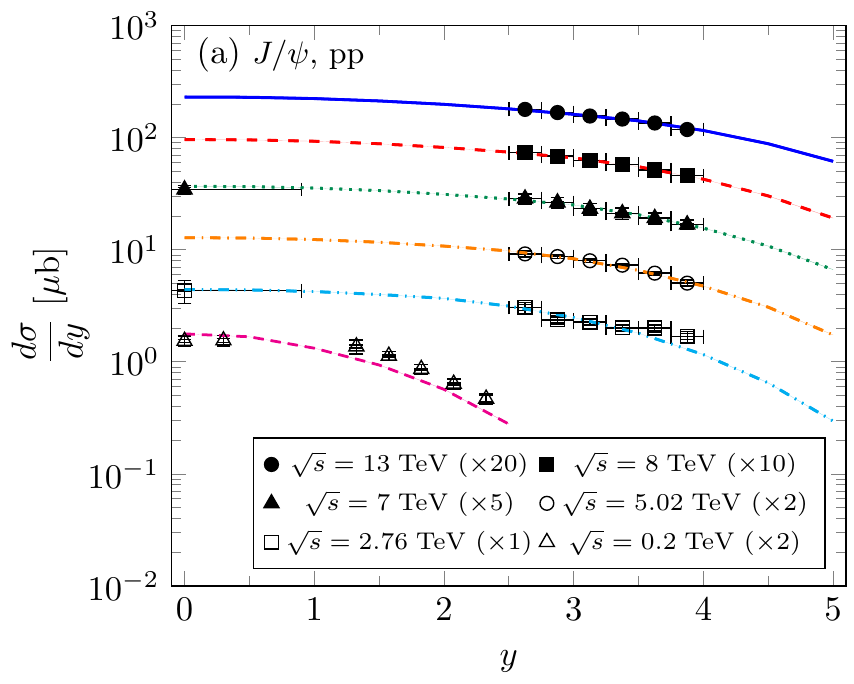}
\includegraphics[width=0.45\linewidth]{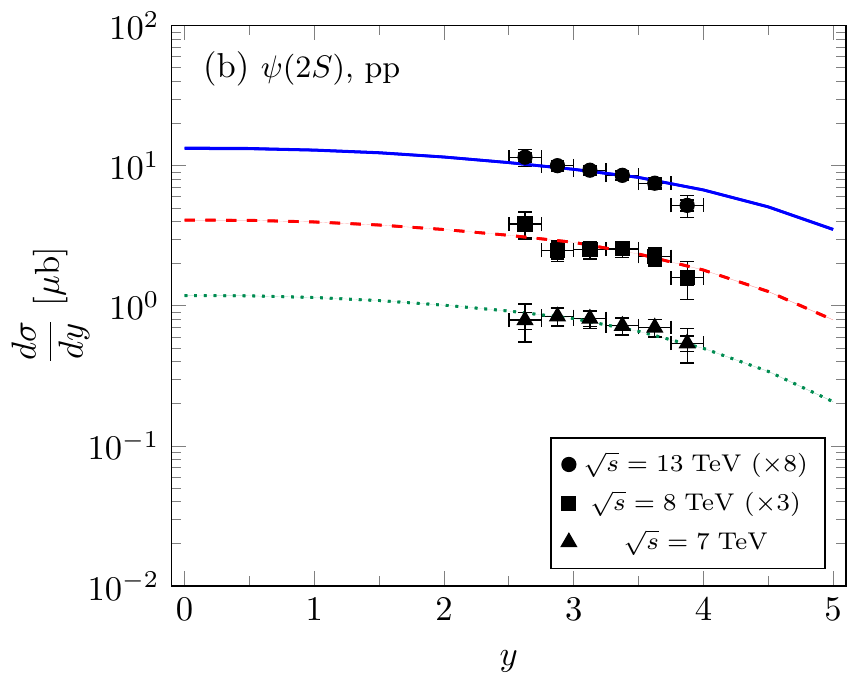}
	\caption{(Color online)
	Rapidity distributions of $J/\psi$ and $\psi(2S)$  in p+p collisions at RHIC and LHC. The various lines correspond to the results for the different values of $\sqrt{s}$. The uncertainty bands reflect the quark mass dependence. Data are taken from Refs.~\cite{Adare:2011vq,Aamodt:2011gj,Abelev:2012kr,Abelev:2014qha,Adam:2015rta,Acharya:2017hjh}.
	}
	\label{fig:Jpsi-Psi2S-y-distribution-pp}
\end{figure}

\begin{figure}
	\centering
	\includegraphics[width=0.45\linewidth]{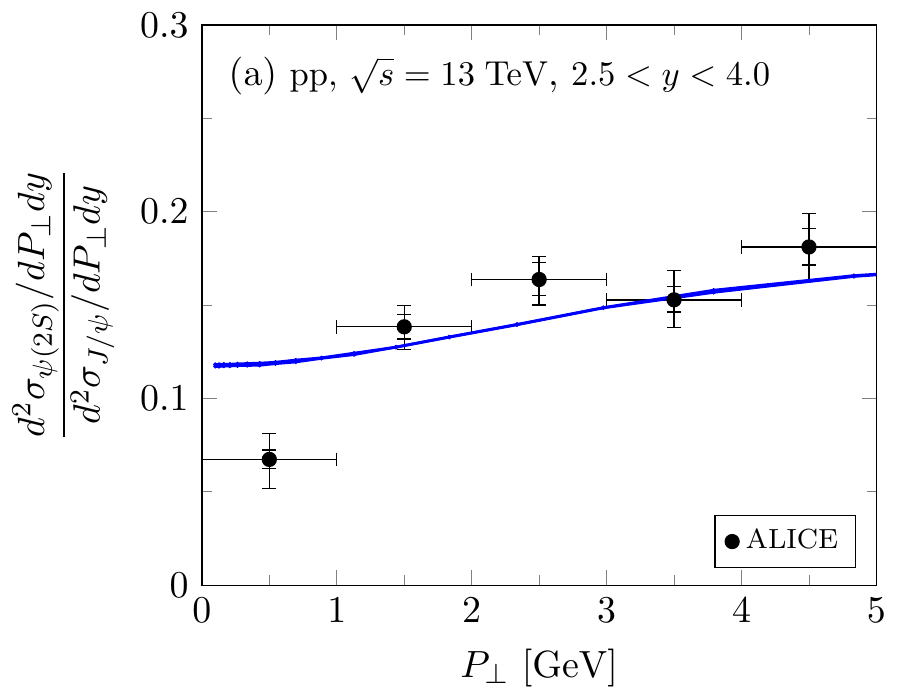}
	\includegraphics[width=0.45\linewidth]{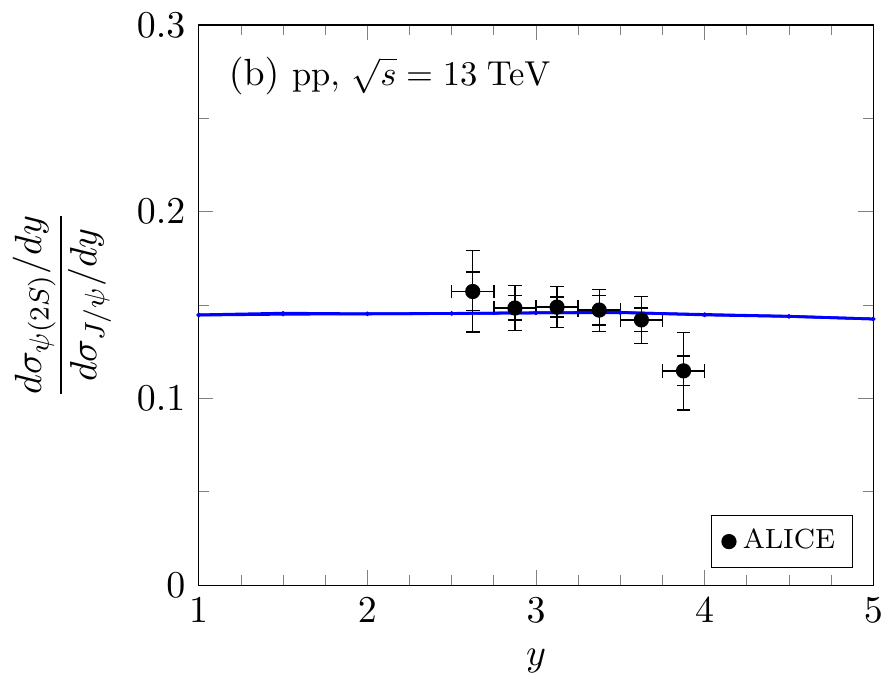}
	\caption{(Color online)
	Ratios of the differential cross-section for $J/\psi$ production in p+p collisions at $\sqrt{s}=13$\;TeV relative to that for $\psi(2S)$ production. The uncertainty bands reflect the quark mass dependence. Data from Refs.~\cite{Acharya:2017hjh}.
	}
	\label{fig:Psi2S-Jpsi-xsection-ratio-lhc}
\end{figure}

We shall now compare our results in the CGC+ICEM framework to data in p+p collisions from RHIC and the LHC. Figure~\ref{fig:Jpsi-Psi2S-pt-distribution-pp} shows a  comparison of this model to the $P_\perp$ spectra of $J/\psi$ and $\psi(2S)$ at various scattering energies and rapidities from RHIC to the LHC. Each of the uncertainty bands reflects the quark mass dependence: $m=(1.3\!\!-\!\!1.4)$\;GeV, although the widths are actually small. The transition probability $F_{\psi}$ is determined by fitting it to data at each $\sqrt{s}$ and $y$ by minimizing the $\chi^2$. The model compares well to the data albeit at RHIC, deviations are seen above $P_\perp=3.5$ GeV for the
$J/\psi$ data. The model agrees well with the data for $\psi(2S)$, within the significant error bars in the data.

In Fig.~\ref{fig:Jpsi-Psi2S-y-distribution-pp}, we show the rapidity ($y$) distributions of $J/\psi$ and $\psi(2S)$ production in p+p collisions at RHIC and the LHC. The cross-sections are obtained by integrating the differential cross sections over the entire $P_\perp$ range up to $P_\perp=10\;{\rm GeV}$. The nonperturbative parameter $F_{\psi}$ in the ICEM is determined by fitting the data at each $\sqrt{s}$ and $y$. The results for these are presented in tabular form and discussed at length in the Appendix.

In Fig.~\ref{fig:Psi2S-Jpsi-xsection-ratio-lhc}, we plot the ratios of the differential cross-sections for $\psi(2S)$ production in p+p collisions with those for $J/\psi$ production in our framework and compare these to LHC data at $\sqrt{s}=13\;{\rm TeV}$. The $P_\perp$ distribution of the ratio in our CGC+ICEM framework agrees nicely with the data within experimental uncertainties. We have confirmed that the numerical results in the CGC+ICEM can also predict the ratios at the LHC $\sqrt{s}=8,\;7\;{\rm TeV}$ and RHIC $\sqrt{s}=0.2\;{\rm TeV}$. We do not show those results here because their $P_\perp$ distributions for the ratios are very similar to the one in Fig.~\ref{fig:Psi2S-Jpsi-xsection-ratio-lhc}. We also observe in the right figure that the rapidity distribution is reproduced, albeit the model overshoots the data slightly at $y=4$.

We would like to comment here on the trend of the $P_\perp$ dependence of the ratio shown in Fig.~\ref{fig:Psi2S-Jpsi-xsection-ratio-lhc}. In leading order kinematics, Eq.~(\ref{eq:lo-kinematics}) as employed in the conventional CEM indicates that $x_2^{J/\psi}\lesssim x_2^{\psi(2S)}$ within the mass range $m_{\psi}<M<2m_D$. The modification in the ICEM of the transverse momentum of the $c\bar c$ pair as $P^\prime_\perp=(M/m_\psi)P_\perp$ through the hadronization process makes the increase of $x_2$ for $\psi(2S)$ slower than that of the $J/\psi$ as $P_\perp$ becomes larger. As a result, the ratio of $\psi(2S)$ to the $J/\psi$ production cross-section increases when $P_\perp$ is large. Thus this improved CGC+ICEM reproduces the trend of the data correctly.
The conventional CEM consistently predicts that the ratio of $\psi(2S)$ to $J/\psi$ production is a  constant fully determined by $F_{J/\psi}$ and $F_{\psi(2S)}$.

\subsection{Results for p+A collisions}

\begin{figure}
	\centering
	\includegraphics[width=0.45\linewidth]{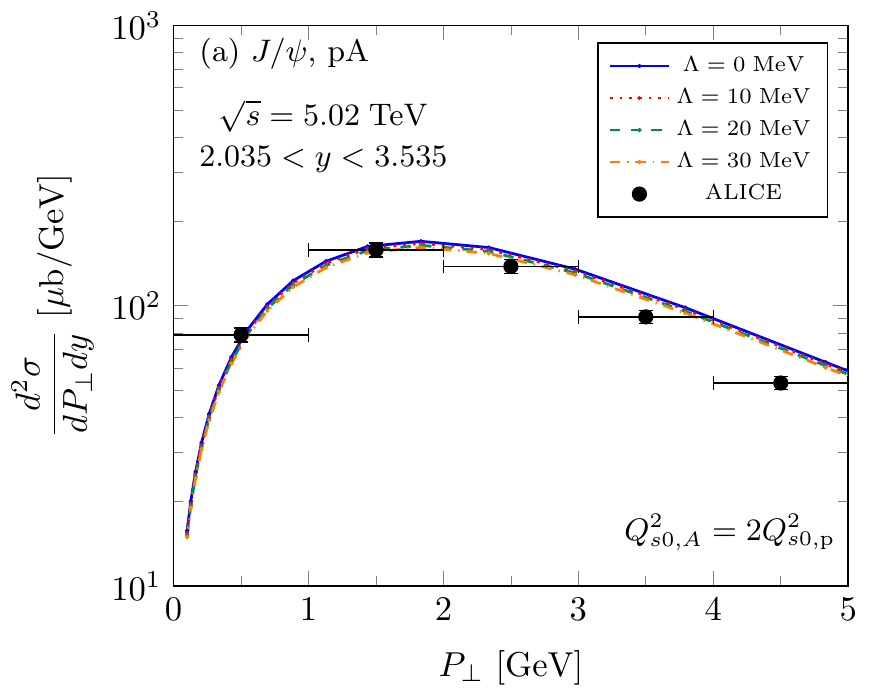}
	\includegraphics[width=0.45\linewidth]{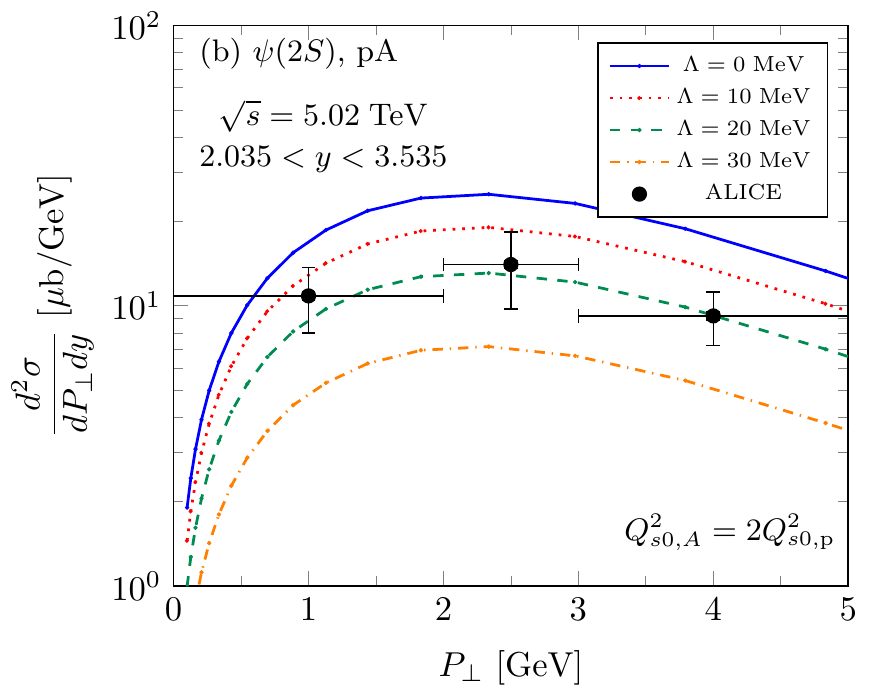}
	\caption{(Color online)
	$P_\perp$ distribution of forward $J/\psi$ and $\psi(2S)$ production in p+A collisions at the LHC for varying values of  $\Lambda$. The quark mass is fixed as $m=1.3\;{\rm GeV}$. The initial saturation scale for the target nucleus is chosen to be  $Q_{s0,A}^2=2\,Q_{s0,{\rm p}}^2$. Data are taken from Refs.~\cite{Abelev:2014zpa,Adam:2015iga}.
	}
	\label{fig:Jpsi-Psi2S-pt-distribution-lhc}
\end{figure}

\begin{figure}
	\centering
	\includegraphics[width=0.45\linewidth]{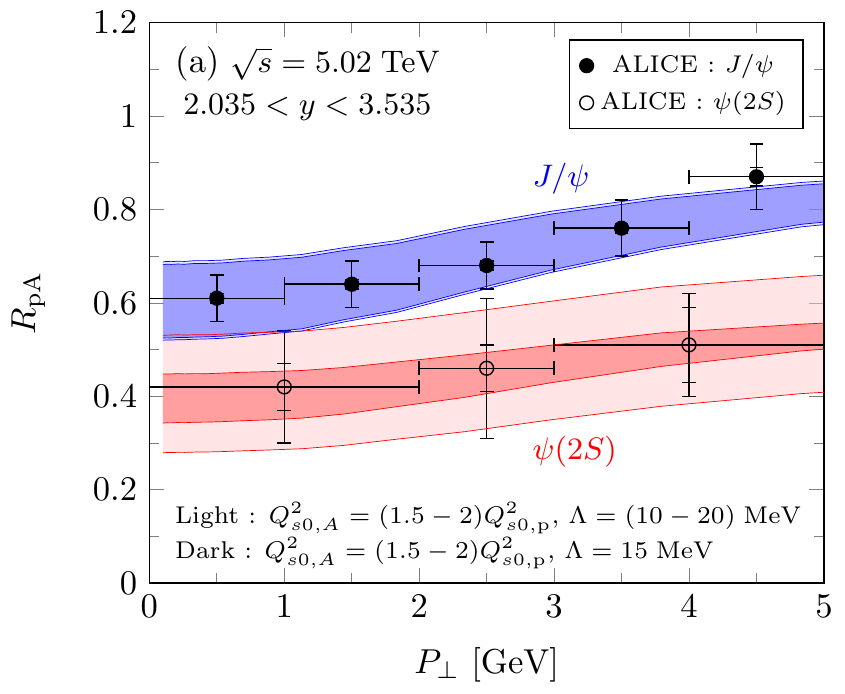}
	\includegraphics[width=0.45\linewidth]{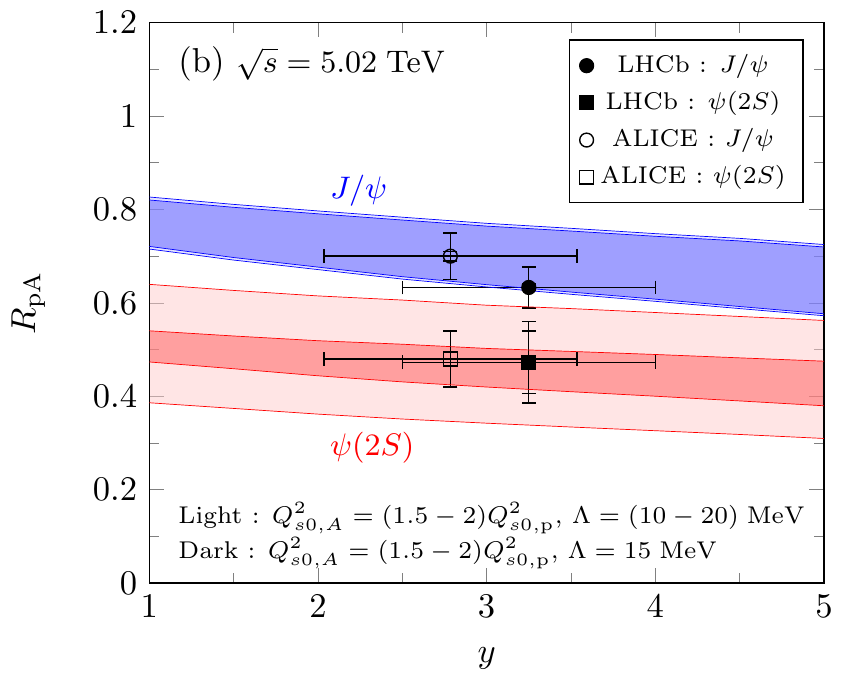}
	\caption{(Color online)
	(a) Nuclear modification factors of $J/\psi$ and $\psi(2S)$ vs $P_\perp$. Data are taken from Refs.~\cite{Abelev:2014zpa,Adam:2015iga}. (b) $y$-dependence of nuclear modification factors of $J/\psi$ and $\psi(2S)$.  Data are taken from Refs.~\cite{Abelev:2014zpa,Aaij:2016eyl}. The dark shaded uncertainty band corresponds to a fixed value of $\Lambda=15$ MeV but varying $Q_{s0,A}^2=(1.5\!\!-\!\!2.0)Q_{s0,{\rm p}}^2$. The light shaded uncertainty bands display the further uncertainties in varying $\Lambda$ in the range $\Lambda=10\!\!-\!\!20$\;MeV on top of the variations in the choice of $Q_{s0,A}^2$.
	}
	\label{fig:LHC-RpA}
\end{figure}

\begin{figure}
	\centering
	\includegraphics[width=0.45\linewidth]{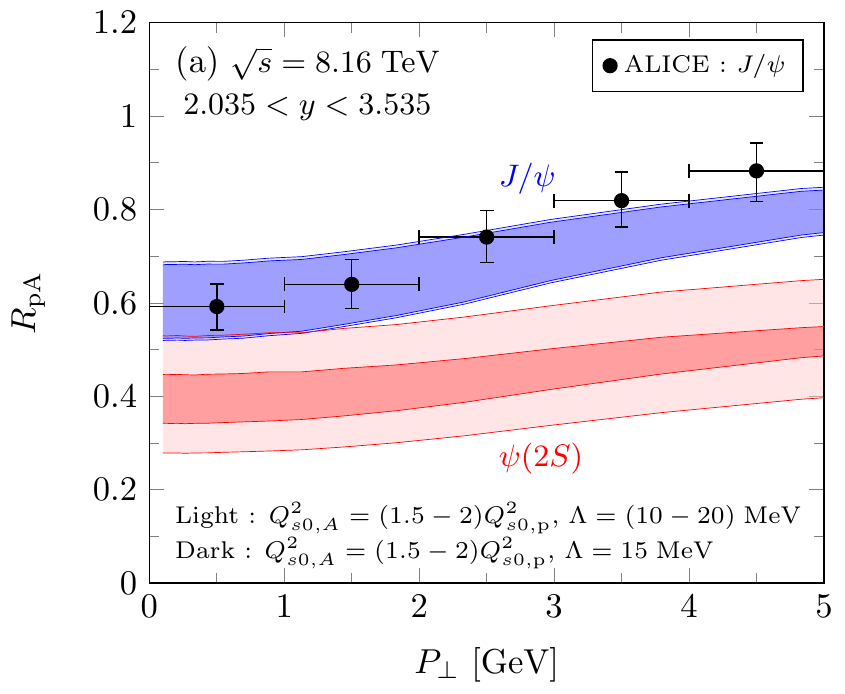}
	\includegraphics[width=0.45\linewidth]{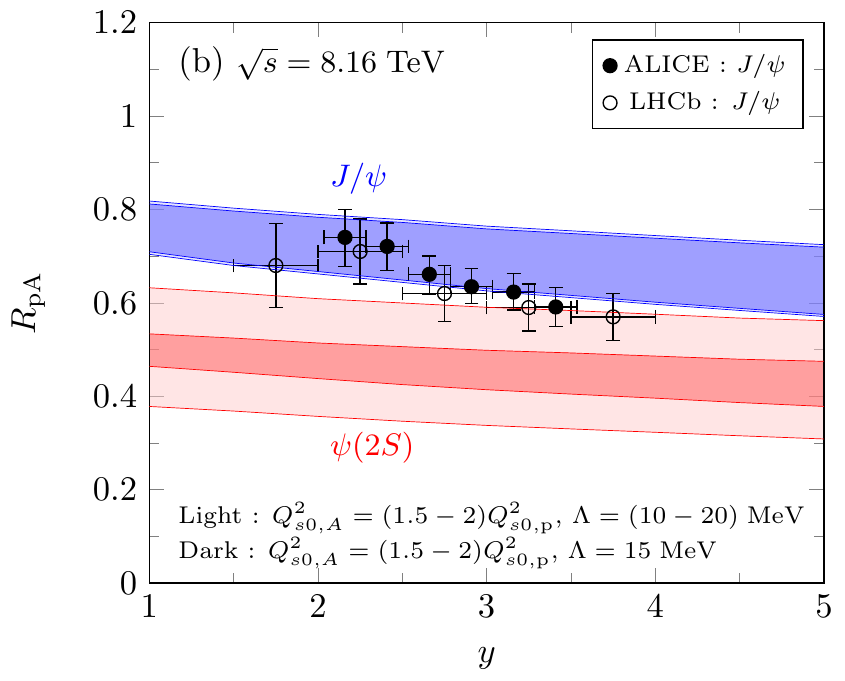}
	\caption{(Color online)
	Comparison to $J/\psi$ data and predictions for $\psi(2S)$ data at the LHC for $\sqrt{s}=8.16$\;TeV/nucleon. Notations are the same as Fig.~\ref{fig:LHC-RpA}. $\Lambda$ is fixed to be the same at both $\sqrt{s}=5.02\;{\rm TeV}$ and $\sqrt{s}=8.16\;{\rm TeV}$. ALICE preliminary data are taken from Ref.~\cite{ALICEprelim2017}. LHCb data are taken from Ref.~\cite{Aaij:2017cqq}.
	}
	\label{fig:LHC-prediction}
\end{figure}

\begin{figure}
	\centering
	\includegraphics[width=0.45\linewidth]{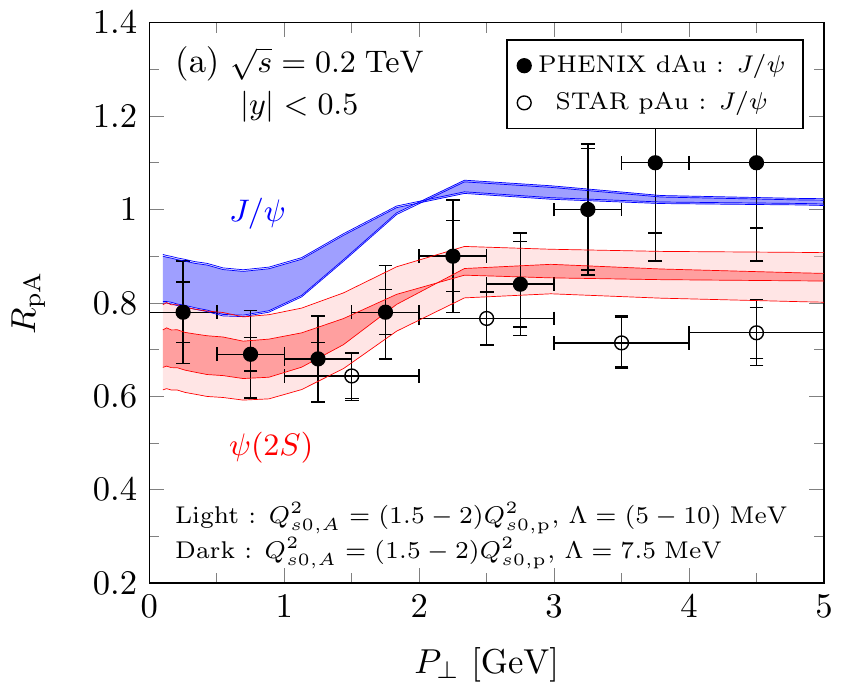}
	\includegraphics[width=0.45\linewidth]{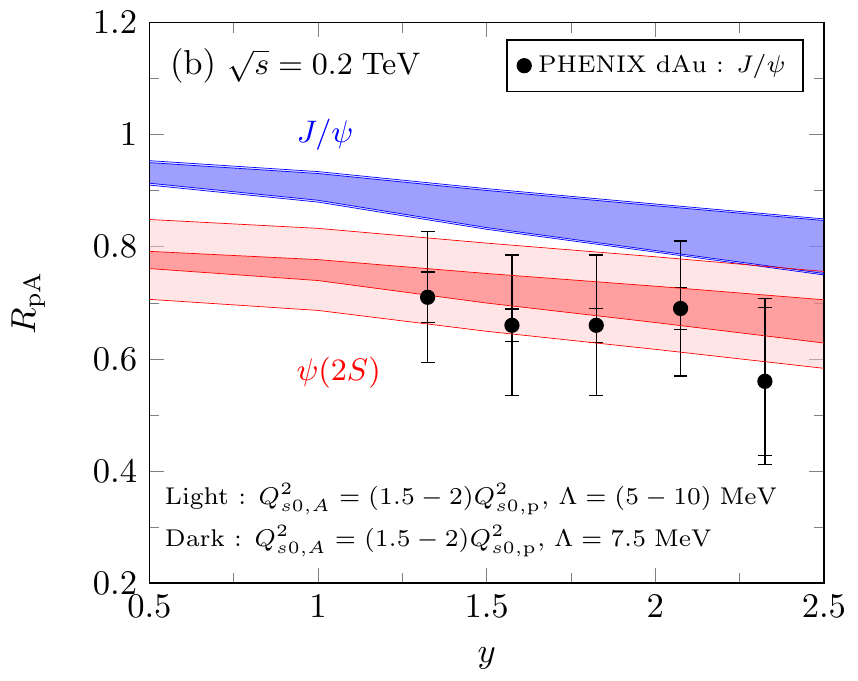}	
	\caption{(Color online)
	Comparison to $J/\psi$ data and predictions for $\psi(2S)$ data at RHIC for $\sqrt{s}=0.2$\;TeV/nucleon. Notations are the same as Fig.~\ref{fig:LHC-RpA} but here $\Lambda=7.5$\;MeV is the fixed valued for the dark shaded band. The  light bands reflect the variation in $\Lambda=5\!\!-\!\!10$\;MeV. Data on $J/\psi$ production in d+Au collisions are taken from Refs.~\cite{Adare:2010fn,Adare:2012qf}. STAR preliminary data are from Ref.~\cite{Todoroki:2017ngs}.
	}
	\label{fig:RHIC-prediction}
\end{figure}

\begin{figure}
	\centering
	\includegraphics[width=0.45\linewidth]{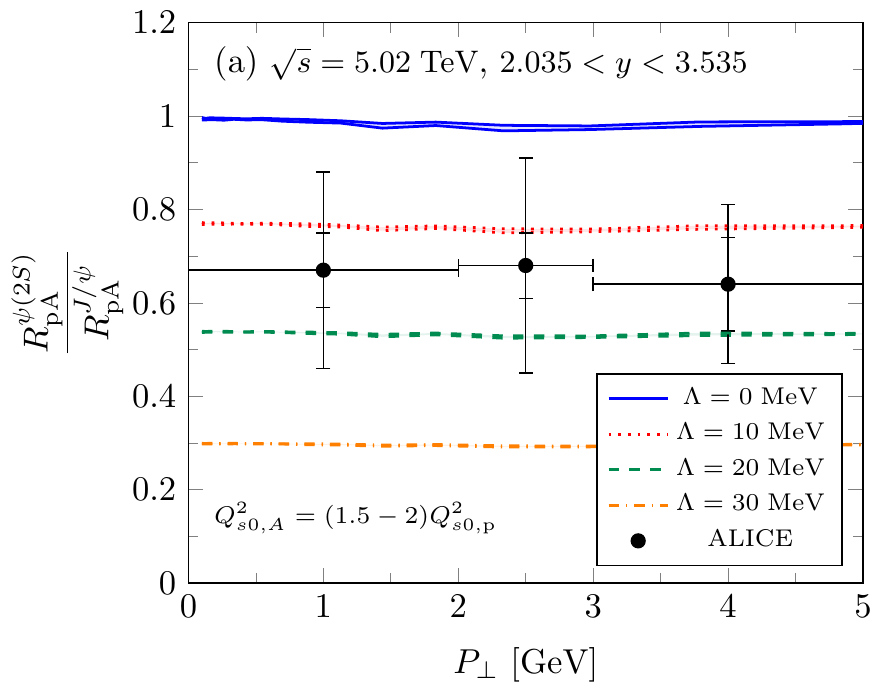}
	\includegraphics[width=0.45\linewidth]{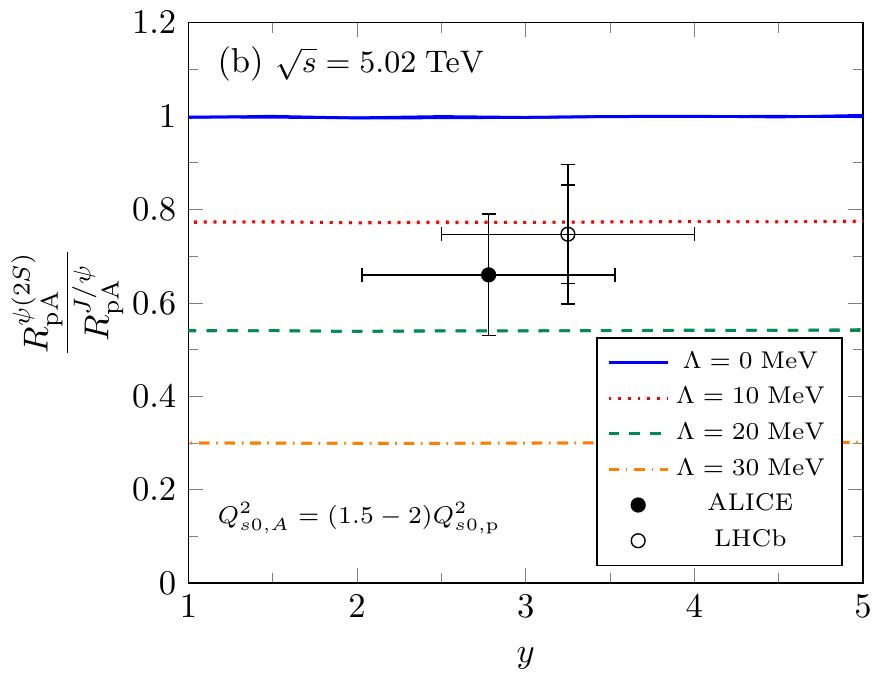}
	\caption{(Color online)
	(a) $P_\perp$ dependence and (b) rapidity dependence of ratios of $R_{\rm{pA}}$ of $J/\psi$ relative to that of $\psi(2S)$ at the LHC obtained by varying the shown values of $\Lambda$. Uncertainty bands reflect the dependence of the model on the initial saturation scale.
	Data are taken from Refs.~\cite{Abelev:2014zpa,Aaij:2016eyl}.
	}
	\label{fig:LHC-RpA-ratio}
\end{figure}

\begin{figure}
	\centering
	\includegraphics[width=0.45\linewidth]{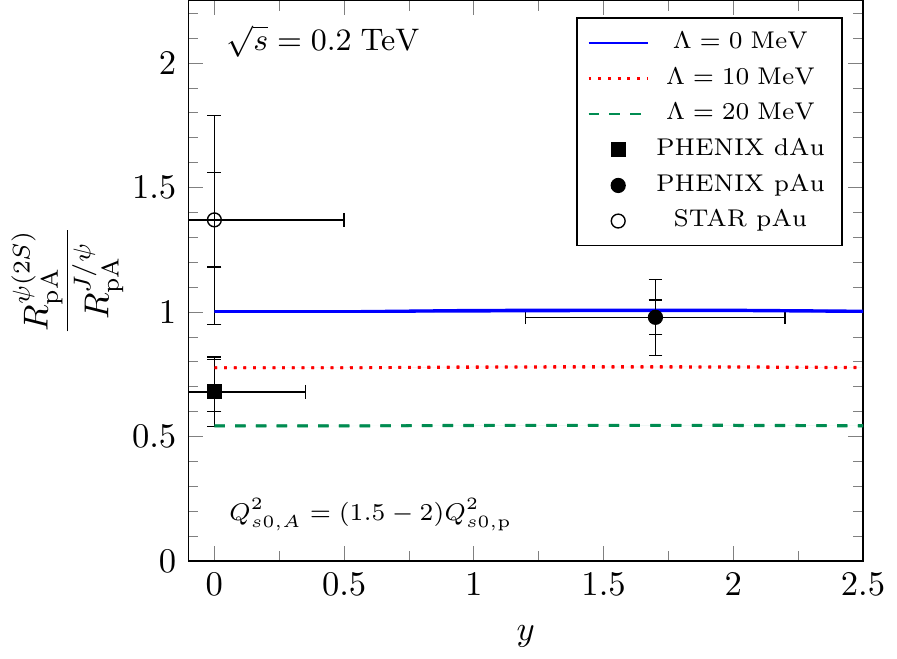}
	\caption{(Color online)
	Ratios of $R_{\rm{pA}}$ for $J/\psi$ relative to the $R_{\rm{pA}}$ for $\psi(2S)$ at RHIC. Notations are the same as in  Fig.~\ref{fig:LHC-RpA-ratio}.
	PHENIX data are taken from Refs.~\cite{Adare:2013ezl,Adare:2016psx}. STAR preliminary data are available in Ref.~\cite{Todoroki:2017ngs}.
	}
	\label{fig:RHIC-RpA-ratio}
\end{figure}

We will now compare our model to data on Onium production in p+A collisions. In order to discuss the $P_\perp$ spectra of $J/\psi$ and $\psi(2S)$ productions in p+A collisions, we will need to first determine the effective transverse area of the target nucleus. Our naive expectation is that nuclear modification factor $R_{\rm pA}$ for Onium production should approach unity at asymptotically high $P_\perp$ because coherent interactions of the produced Onia with the target nucleus should be negligible at these values of $P_\perp$. The nuclear modification factor is defined to be
\begin{align}
R_{\rm pA}=\dfrac{1}{A}\dfrac{d^3\sigma_{\rm pA}/d^2P_\perp dy}{d^3\sigma_{\rm pp}/d^2P_\perp dy}.
\end{align}
Our asymptotic condition then leads to~\cite{Ma:2015sia}
\begin{align}
R_{\rm pA}\stackrel{{\rm high}\;P_\perp}{\approx}\frac{1}{A}\frac{\pi R_A^2}{\pi R_{\rm p}^2}\frac{Q_{s,A}^{2\gamma}}{Q_{s,{\rm p}}^{2\gamma}}\approx \frac{1}{A}\frac{\pi R_A^2}{\pi R_{\rm p}^2}\frac{Q_{s0,A}^{2\gamma}}{Q_{s0,{\rm p}}^{2\gamma}}\stackrel{P_\perp\rightarrow\infty}{\longrightarrow}1 \,.
\label{eq:RA}
\end{align}
The MV model parametrization gives $\gamma=1$. As a result,  $R_A$ can be determined uniquely as $R_A=\sqrt{A/N} R_{\rm p}$ with $N=Q_{s0,A}^2/Q_{s0,{\rm p}}^2$. One should keep in mind that the effective radius $R_A$ is no other than the normalization parameter to obtain $R_{{\rm p}A}=1$ at high $P_\perp$.

Figure~\ref{fig:Jpsi-Psi2S-pt-distribution-lhc} shows the numerical results for the differential cross-section for $J/\psi$ and $\psi(2S)$ production in p+A collisions at the LHC. For $F_{J/\psi}$ and $F_{\psi(2S)}$, we have used the averaged numerical values obtained by fitting data in p+p collisions at $\sqrt{s}=13,\;8,\;7\;{\rm TeV}$ in the rapidity range $2.5<y<4.0$. As shown in the Appendix, we  found that the numerical values of $F_{J/\psi}$ are only weakly dependent on the center-of-mass energy.

The CGC+ICEM  can describe the differential cross section for $J/\psi$ production at low $P_\perp$ up to nearly $P_\perp=4\;{\rm GeV}$. They key features of the comparison in Fig.~\ref{fig:Jpsi-Psi2S-pt-distribution-lhc} are as follows:  the $P_\perp$ distribution of $J/\psi$ production in p+A collisions is nearly identical to the $\Lambda=0$\;MeV when $\Lambda$ is varied over the range shown. In contrast, the additional soft color exchanges in p+A collisions significantly  affect the $P_\perp$ spectrum of $\psi(2S)$. This occurs even though the $\Lambda$ values shown are very soft when compared to $\Lambda_{\rm QCD}$. The plots in Fig.~\ref{fig:Jpsi-Psi2S-pt-distribution-lhc} indicate that the best fits are obtained for $\Lambda=(10\!\!-\!\!20)$\;MeV. As we observed previously, the dependence of the $\psi(2S)$ cross-section on $\Lambda$ reflects simply the ease with which additional soft color exchanges in p+A collisions can break up the $\psi(2S)$ by providing the energy to push the bound Onia over the $D\bar{D}$ threshold. In contrast, these soft color exchanges have no visible impact on the $J/\psi$ since it is relatively far more strongly bound.

The importance of soft color exchanges from comovers is more pronounced in the nuclear modification factor. The $P_\perp$ and $y$ distributions of $R_{\rm pA}$ for $J/\psi$ and $\psi(2S)$ at the LHC in the forward rapidity region are illustrated in Fig.~\ref{fig:LHC-RpA}.  We employ here the same values of $\Lambda$. The CGC+ICEM framework describes the $R_{\rm pA}$ for $J/\psi$ and $\psi(2S)$ nicely. Without the comover interaction at $\Lambda=0$\;MeV, the $R_{\rm pA}$ of $\psi(2S)$ is almost the same as that of $J/\psi$. Figure~\ref{fig:LHC-prediction} displays our comparison to the $R_{\rm pA}$ for $J/\psi$ and our prediction for $\psi(2S)$ at $\sqrt{s}=8.16$\;TeV. Here we have used the same values of $\Lambda$ as at $\sqrt{s}=5.02\;{\rm TeV}$. Results for $R_{\rm pA}$ at RHIC are shown in Fig.~\ref{fig:RHIC-prediction}.  Our curve for $J/\psi$ is slightly above the PHENIX data in dAu collisions and the STAR preliminary data in pAu collisions at low $P_\perp$. Nevertheless, the experimental uncertainties are large. Our results shows a Cronin peak around $P_\perp\sim2$\;GeV and a weaker $J/\psi$ suppression than the LHC. This is because the multiple scattering effect in the target nucleus without rapidity evolution is only accounted for in the mid rapidity region at RHIC where $x_{1,2}>0.01$. Even for the small values of $\Lambda$ shown in the figure, one obtains a stronger suppression of $\psi(2S)$ relative to the $J/\psi$.

Figure~\ref{fig:LHC-RpA-ratio} contains the ratio of ratios --- between the $R_{\rm pA}$ of $\psi(2S)$ and that of the $J/\psi$. The CGC+ICEM prediction is that the $P_\perp$ distribution and the rapidity distribution of the double ratio is nearly flat; this is because we have assumed that $\Lambda$ is independent of $P_\perp$ and $y$. An advantage of this double ratio is that one expects many systematic uncertainties to cancel. In particular, the uncertainties related to the initial saturation scale for the target nucleus and that of the quark mass are reduced significantly. On the other hand, we observe that the double ratios are clearly suppressed as the value of $\Lambda$ increases up to values in the range $\Lambda=(10\!\!-\!\!20)\;{\rm MeV}$, although the data have large uncertainties. Nevertheless, these plots show very clearly that the suppression of the double ratio can be controlled by $\Lambda$ alone. Figure~\ref{fig:RHIC-RpA-ratio} similarly shows the double ratio at RHIC. At present, the RHIC data does not show a strong suppression; the statistical uncertainties are however large. Our lower bound of $\Lambda =10$\;MeV is compatible, within these large errors, with the data from PHENIX at both central and forward rapidities.

\section{Summary}

We studied in this paper $J/\psi$ and $\psi(2S)$ production in p+p and p+A collisions at RHIC and the LHC in the CGC+ ICEM framework.
The short distance cross-section depends on the convolution of the transverse momentum $k_\perp$ dependent gluon distribution for the projectile proton and  $k_\perp$ dependent multipoint Wilson line correlators in the target. Small-$x$ evolution effects are accounted for via the running coupling BK equation. The ICEM  parametrizes soft gluons exchanges between the $c\bar c$ and color sources as well as soft gluon emissions from the $c\bar c$.  We find that this CGC+ICEM framework provides a good description of the differential cross-sections for both $J/\psi$ and $\psi(2S)$ production in p+p collisions at low $P_\perp$ at RHIC and at a range of LHC energies. We also observe that the ratio of the differential cross-sections for $J/\psi$ and $\psi(2S)$ is reproduced for a wide energy range.

The surprisingly large suppression of $\psi(2S)$ production relative to that of $J/\psi$ production in p+A collisions at both RHIC and the LHC has widely been interpreted as arising from final state interactions with hadron comovers. We argued here that this large suppression can be explained by factorization breaking soft color exchanges that are enhanced in p+A collisions. We showed that these effects could be implemented heuristically in the ICEM by reducing the upper limit of the invariant mass of the $c\bar{c}$ pair by a parameter $\Lambda$ that represents the momentum kick delivered by the additional soft color exchanges in p+A collisions.  After fitting the p+p data with  $\Lambda=0$ and extracting values of the nonperturbative constant $F_\psi$ for different collision energies, we kept $F_\psi$ fixed for the p+A collisions and studied the dependence of p+A results on $\Lambda$. We find excellent fits of data from the LHC  for $\Lambda\approx (10\!\!-\!\!20)$\;MeV at $\sqrt{s}=5.02$\;TeV and for $\Lambda\lesssim 10$\;MeV at RHIC for  $\sqrt{s}=0.2$\;TeV. These values of $\Lambda$ are much smaller than $\Lambda_{\rm QCD}$.
Our results therefore suggest that enhanced soft color exchanges are sufficient to explain the observed pattern of suppression in these collisions.

\begin{acknowledgments}

The authors are grateful to Lijuan Ruan for a discussion on the preliminary STAR data. They would also like to thank Peter Petreczky and Jianwei Qiu for useful discussions. K.W. is grateful to Hirotsugu Fujii for help regarding numerical implementations of the rcBK equation. He is supported by Jefferson Science Associates, LLC under  U.S. DOE Contract No.~DE-AC05-06OR23177 and U.S. DOE Grant No.~DE-FG02-97ER41028. The work of K.W. was also supported by the National Science Foundation of China (NSFC) under Grant No.~11575070. R.V's research is supported by the U.S. Department of Energy, Office of Science, Office of Nuclear Physics, under Contract No.~DE-SC0012704. R.V. would also like to thank the Alexander von Humboldt Foundation and ITP Heidelberg for support, and ITP Heidelberg for their kind hospitality during the completion of this work.

\end{acknowledgments}

\appendix

\section{Tabular values for $F_\psi$ from fits to p+p data}

We tabulate here in Tables I-IV the values of $F_\psi$ extracted from fits to the $P_\perp$ and rapidity distributions of $J/\psi$ and $\psi(2S)$ in p+p collisions at RHIC and the LHC
\footnote{In this paper, the standard unweighted $\chi$-squared minimization is employed simply to determine the overall factors $F_\psi$. An explicit expression of fitted $F_\psi$ can be written as
\begin{align}
F_\psi=\frac{\sum_{i=1}^n y_i f(x_i)}{\sum_{i=1}^n f(x_i)^2}\pm \hat{\sigma}_F
\end{align}
where $\hat\sigma_F=\sqrt{\sigma^2/\sum_{i=1}^n f(x_i)^2}$ is the unweighted deviation of the fit-parameter $F_\psi$ with $\sigma^2=\frac{1}{n-1}\sum_{i=1}^n[y_i-F_\psi f(x_i)]^2$. $y_i$ are data points of a sample at point $x_i$. $n$ is the number of data points which we consider in parameter fitting. $f(x_i)$ are corresponding theoretical results, $d\sigma_\psi$ except for $F_{\psi}$. In this paper, $\chi^2$ is evaluated as
\begin{align}
\chi^2=\frac{1}{\sigma_{\rm err}^2}\sum_{i=1}^n[y_i-F_\psi f(x_i)]^2,
\end{align}
where $\sigma_{\rm err}^2=\frac{1}{n}\sum_{i=1}^n (y_i^{\rm err})^2$ being the variance of the data. $y_i^{\rm err}$ includes statistical error and uncorrelated systematic error at point $x_i$.
}.
In Table I, we show the values extracted from the fits to the $P_\perp$ distribution in Fig.~\ref{fig:Jpsi-Psi2S-pt-distribution-pp} below $P_\perp=6$\;GeV. The results are shown for two values of the quark mass, $m=1.3$ GeV and $m=1.4$ GeV. At the LHC, the central values of $F_{J/\psi}$ are about 20\% larger for $m=1.4$ GeV than at $m=1.3$ GeV. The variation for each $m$ as a function of energy from $2.76$ TeV to $13$ TeV is only  by at most $\sim 10$\%. At RHIC energies, the central values of $F_{J/\psi}$ are somewhat larger, being greater by about $30$\%.

The corresponding values for $F_{\psi(2S)}$ are shown in Table II. These are approximately a factor of 2 larger than $F_{J/\psi}$. The central values at the LHC energies are quite stable, but are about $40$\% smaller than those at the RHIC energies. The quark mass dependence is weaker here than for $J/\psi$.

The fits to some of the LHC data on the $P_\perp$ distribution of $J/\psi$ provide large values of $\chi^2/d.o.f.(\gg 1)$ which signify in general that the fits are poor at the LHC. Nevertheless, we can control the overall factor only and our fitting method indeed determines the reasonable values of $F_{J/\psi}$. Meanwhile, for $\psi(2S)$ production, the fits to the LHC data provide slightly better values of $\chi^2/d.o.f.$ compared to the fits for $J/\psi$ production.

A similar pattern is seen for $F_\psi$ extracted from the rapidity distributions in Fig.~\ref{fig:Jpsi-Psi2S-y-distribution-pp}. Given the variation in energies studied, the results for $F_\psi$ are remarkably stable with small values of $\chi^2/d.o.f.$ except for RHIC energy.

The numerical values of $F_{J/\psi}$ and $F_{\psi(2S)}$ extracted from $d\sigma/d^2P_\perp dy$ in p+p collisions must be universal and applied to p+A collisions within the same rapidity range. However, rapidity in the center-of-mass frame in p+A collision at the LHC is shifted by $0.465$ from that in the laboratory frame.  We assume that $F_{J/\psi}$ and $F_{\psi(2S)}$ remain the same at both $2.5<y<4.0$ and $2.035<y<3.535$ in the LHC energies. For $J/\psi$ production with $m=1.3$\;GeV, the averaged numerical value of $F_{J/\psi}$ obtained from those at $\sqrt{s}=13,\;8,\;7\;{\rm TeV}$ in the rapidity range $2.5<y<4.0$ is $0.215\pm 1.36\times10^{-2}$. This value is consistent with those at $\sqrt{s}=5.02,\;2.76$\;TeV within the errors. Therefore, this averaged value of $F_{J/\psi}$ can be used to evaluate the differential cross section in p+A collisions in Fig.~\ref{fig:Jpsi-Psi2S-pt-distribution-lhc}. Likewise, for $\psi(2S)$ with $m=1.3$\;GeV, the fit values of $F_{\psi(2S)}$ at $\sqrt{s}=13,\;8,\;7\;{\rm TeV}$ in the rapidity range $2.5<y<4.0$ is $0.509\pm3.49\times10^{-2}$, which is used in p+A collisions at $\sqrt{s}=5.02$\;TeV.

\begin{table}[tbp]
\renewcommand\arraystretch{1.25}
\begin{center}
\caption{Fitted values and errors of $F_\psi$ for $d\sigma/d^2P_\perp dy$ of $J/\psi$ production in p+p collisions. Numbers in the bracket next to $F_{J/\psi}$ represent the heavy quark mass value: $m=1.3\;{\rm GeV}$ or $1.4\;{\rm GeV}$.}
\begin{tabular}{ccc|cc|cc}
\hline
\hline
\hspace{0.01\textwidth} $\sqrt{s}$ [TeV] \hspace{0.01\textwidth} & \hspace{0.04\textwidth} $y$ bin \hspace{0.04\textwidth} & Data points \hspace{0.075\textwidth}& \hspace{0.05\textwidth} $F_{J/\psi}$ (1.3) \hspace{0.075\textwidth} &  $\chi^2/d.o.f.$ \hspace{0.075\textwidth}&   \hspace{0.075\textwidth}$F_{J/\psi}$ (1.4) \hspace{0.05\textwidth}   &  $\chi^2/d.o.f.$  \hspace{0.075\textwidth}\\
\hline
13   & $2.5<y<4.0$ &  7   &     0.222 $\pm$ 1.41$\times10^{-2}$ & 8.7     &   0.262 $\pm$ 1.72$\times10^{-2}$  & 9.3  \\
8     & $2.5<y<4.0$ &  6   &    0.231 $\pm$ 1.50$\times10^{-2}$ & 4.3      &   0.273 $\pm$ 1.86$\times10^{-2}$   &  4.8 \\
7     & $|y|<0.9$      &  5   &    0.178 $\pm$ 9.21$\times10^{-3}$  & 0.33   &   0.211 $\pm$  1.12$\times10^{-2}$   &  0.35 \\
7     & $2.5<y<4.0$ & 6    &    0.192 $\pm$ 1.19$\times10^{-2}$ & 5.3      &   0.228  $\pm$ 1.49$\times10^{-2}$   & 5.9  \\
5.02& $2.5<y<4.0$ &  6   &    0.207 $\pm$  1.32$\times10^{-2}$   & 7.0   &   0.247 $\pm$  1.66$\times10^{-2}$   & 7.8  \\
2.76& $2.5<y<4.0$ &  6   &    0.208 $\pm$  7.27$\times10^{-3}$  & 0.88  &   0.249 $\pm$  9.58$\times10^{-3}$   & 1.1  \\
0.2  & $|y|<0.35$    &  21  &    0.251 $\pm$  7.64$\times10^{-3}$  & 0.93 &   0.314 $\pm$   9.14$\times10^{-3}$   & 0.85  \\
0.2  & $1.2<y<2.4$ &  24  &    0.417 $\pm$  5.13$\times10^{-3}$  & 0.35  &   0.516 $\pm$  7.16$\times10^{-3}$   & 0.44  \\
\hline
\hline
\end{tabular}
\end{center}
\label{table1}
\end{table}
\begin{table}[tbp]
\renewcommand\arraystretch{1.25}
\begin{center}
\caption{Fitted values and errors of $F_\psi$ for $d\sigma/d^2P_\perp dy$ of $\psi(2S)$ production in p+p collisions.}
\begin{tabular}{ccc|cc|cc}
\hline
\hline
\hspace{0.01\textwidth} $\sqrt{s}$ [TeV] \hspace{0.01\textwidth} & \hspace{0.04\textwidth} $y$ bin \hspace{0.04\textwidth} & Data points \hspace{0.075\textwidth}& \hspace{0.05\textwidth} $F_{\psi(2S)}$ (1.3) \hspace{0.075\textwidth} & $\chi^2/d.o.f.$ \hspace{0.075\textwidth}&  \hspace{0.075\textwidth}$F_{\psi(2S)}$ (1.4) \hspace{0.05\textwidth} & $\chi^2/d.o.f.$ \hspace{0.075\textwidth}\\
\hline
13 & $2.5<y<4.0$ & 6 & 0.523 $\pm$  3.91$\times10^{-2}$ & 2.2 &   0.589 $\pm$ 4.55$\times10^{-2}$  & 2.4 \\
8 & $2.5<y<4.0$ & 6  & 0.494 $\pm$  1.89$\times10^{-2}$ & 0.22 &  0.562 $\pm$ 2.15$\times10^{-2}$  & 0.22\\
7 & $2.5<y<4.0$ & 6  & 0.511 $\pm$  4.67$\times10^{-2}$ & 1.3 &    0.581 $\pm$  5.45$\times10^{-2}$  & 1.4 \\
0.2 & $|y|<0.35$ & 4  & 0.662 $\pm$  5.89$\times10^{-2}$ & 0.23 &  0.784 $\pm$  6.64$\times10^{-2}$  & 0.21\\
\hline
\hline
\end{tabular}
\end{center}
\label{table2}
\end{table}
\begin{table}[tbp]
\renewcommand\arraystretch{1.25}
\begin{center}
\caption{Fitted values and errors of $F_\psi$ for $d\sigma/dy$ of $J/\psi$ production in p+p collisions.}
\begin{tabular}{cc|cc|cc}
\hline
\hline
\hspace{0.01\textwidth} $\sqrt{s}$ [TeV] \hspace{0.01\textwidth} & Data points \hspace{0.075\textwidth}&  \hspace{0.05\textwidth} $F_{J/\psi}$ (1.3) \hspace{0.075\textwidth} & $\chi^2/d.o.f.$ \hspace{0.075\textwidth}& \hspace{0.075\textwidth}$F_{J/\psi}$ (1.4) \hspace{0.05\textwidth} & $\chi^2/d.o.f.$ \hspace{0.075\textwidth}\\
\hline
13    & 6 & 0.205 $\pm$ 1.32$\times10^{-3}$ & 0.085 & 0.240 $\pm$ 1.55$\times10^{-3}$ & 0.086\\
8      & 6 & 0.223 $\pm$ 9.50$\times10^{-4}$ & 0.017 & 0.262 $\pm$ 1.12$\times10^{-3}$ & 0.017\\
7      & 6 & 0.184 $\pm$ 2.19$\times10^{-3}$ & 0.069 & 0.217 $\pm$ 2.57$\times10^{-3}$ & 0.069\\
5.02 & 6 & 0.193 $\pm$ 2.06$\times10^{-3}$ & 0.17  & 0.228  $\pm$ 2.38$\times10^{-3}$ & 0.16\\
2.76 & 7 & 0.190 $\pm$ 6.79$\times10^{-3}$ & 0.32  & 0.226  $\pm$ 8.00$\times10^{-3}$ & 0.32\\
0.2   & 7 & 0.284 $\pm$ 2.11$\times10^{-2}$ & 2.7    & 0.350  $\pm$ 2.57$\times10^{-2}$ & 2.6\\
\hline
\hline
\end{tabular}
\end{center}
\label{table3}
\end{table}
\begin{table}[tbp]
\renewcommand\arraystretch{1.25}
\begin{center}
\caption{Fitted values and errors of $F_\psi$ for $d\sigma/dy$ of $\psi(2S)$ production in p+p collisions.}
\begin{tabular}{cc|cc|cc}
\hline
\hline
\hspace{0.01\textwidth} $\sqrt{s}$ [TeV] \hspace{0.01\textwidth} & Data points \hspace{0.075\textwidth}&  \hspace{0.05\textwidth} $F_{J/\psi}$ (1.3) \hspace{0.075\textwidth} & $\chi^2/d.o.f.$ \hspace{0.075\textwidth}& \hspace{0.075\textwidth}$F_{J/\psi}$ (1.4) \hspace{0.05\textwidth} & $\chi^2/d.o.f.$ \hspace{0.075\textwidth}\\
\hline
13   & 6 & 0.478  $\pm$ 2.31$\times10^{-2}$  & 0.83 & 0.538 $\pm$ 2.60$\times10^{-2}$ & 0.84\\
8    & 6 & 0.508  $\pm$ 3.40$\times10^{-2}$ & 0.43   &  0.573 $\pm$ 3.84$\times10^{-2}$ & 0.43\\
7    & 6 & 0.474 $\pm$ 1.63$\times10^{-2}$ & 0.12   &   0.535 $\pm$ 1.82$\times10^{-2}$ & 0.12\\
\hline
\hline
\end{tabular}
\end{center}
\label{table4}
\end{table}

\bibliographystyle{utphysMa}
\bibliography{bibTex1.4}

\end{document}